\documentclass[sn-standardnature,iicol]{sn-jnl}

\usepackage{siunitx}
\usepackage{amssymb}
\usepackage{times}
\usepackage{graphicx}
\usepackage{makecell}
\usepackage{rotating}
\usepackage{multibib}
\usepackage{anyfontsize}
\newcites{App}{Methods References}

\graphicspath{{figures/}}

\newcommand{\araa}{Annu.\ Rev.\ Astron.\ Astrophys.} 
\newcommand{\aj}{Astron.\ J.} 
\newcommand{\apj}{Astrophys.\ J.} 
\newcommand{\apjl}{Astrophys.\ J.\ Lett.} 
\newcommand{\apjs}{Astrophys.\ J.\ Suppl.\ Ser.} 
\newcommand{\ao}{Appl.\ Opt.} 
\newcommand{\apss}{Astrophys.\ Space\ Sci.} 
\newcommand{\aap}{Astron.\ Astrophys.} 
\newcommand{\icarus}{Icarus} 
\newcommand{\jgrp}{J.\ Geophys.\ Res.:\ Planets} 
\newcommand{\joss}{J. Open Source Softw.} 
\newcommand{\jqsrt}{J.\ Quant.\ Spectrosc.\ Radiat.\ Transf.} 
\newcommand{\mnras}{Mon.\ Not.\ R.\ Astron.\ Soc.} 
\newcommand{\nat}{Nature} 
\newcommand{\nastro}{Nat.\ Astron.} 
\newcommand{\natas}{Nat.\ Astron.} 
\newcommand{\nmethods}{Nat.\ Methods} 
\newcommand{\pasp}{Publ.\ Astron.\ Soc.\ Pac.} 
\newcommand{\ssr}{Space\ Sci.\ Rev.} 

\jyear{2024}%

\begin{document}

\title{\centering{A High Internal Heat Flux and Large Core in a Warm Neptune Exoplanet}}

\author*[1]{\fnm{Luis} \sur{Welbanks}}\email{luis.welbanks@asu.edu}

\author[2,3]{\fnm{Taylor J.} \sur{Bell}}

\author[4]{\fnm{Thomas G.} \sur{Beatty}}

\author[1]{\fnm{Michael R.} \sur{Line}}

\author[5,6]{\fnm{Kazumasa} \sur{Ohno}}

\author[5]{\fnm{Jonathan J.} \sur{Fortney}}

\author[7]{\fnm{Everett} \sur{Schlawin}}

\author[3]{\fnm{Thomas P.} \sur{Greene}}

\author[8]{\fnm{Emily} \sur{Rauscher}}

\author[9]{\fnm{Peter} \sur{McGill}}

\author[7]{\fnm{Matthew} \sur{Murphy}}

\author[10]{\fnm{Vivien} \sur{Parmentier}}

\author[5]{\fnm{Yao} \sur{Tang}}

\author[2]{\fnm{Isaac} \sur{Edelman}}

\author[5]{\fnm{Sagnick} \sur{Mukherjee}}

\author[1]{\fnm{Lindsey S.} \sur{Wiser}}

\author[11]{\fnm{Pierre-Olivier} \sur{Lagage$^{\text{11}}$}}

\author[11]{\fnm{Achr\`ene} \sur{Dyrek$^{\text{11}}$}}

\author[4]{\fnm{Kenneth E.} \sur{Arnold}}


\affil[1]{\orgdiv{School of Earth and Space Exploration}, \orgname{Arizona State University}, \orgaddress{\city{Tempe}, \state{AZ}, \country{USA}}}

\affil[2]{\orgdiv{Bay Area Environmental Research Institute}, \orgname{NASA's Ames Research Center}, \orgaddress{\city{Moffett Field}, \state{CA}, \country{USA}}}

\affil[3]{\orgdiv{Space Science and Astrobiology Division}, \orgname{NASA's Ames Research Center}, \orgaddress{\city{Moffett Field}, \state{CA}, \country{USA}}}

\affil[4]{\orgdiv{Department of Astronomy}, \orgname{University of Wisconsin-Madison}, \orgaddress{\city{Madison}, \state{WI}, \country{USA}}}

\affil[5]{\orgdiv{Department of Astronomy and Astrophysics}, \orgname{University of California Santa Cruz}, \orgaddress{\city{Santa Cruz}, \state{CA}, \country{USA}}}

\affil[6]{\orgdiv{Division of Science}, \orgname{National Astronomical Observatory of Japan}, \orgaddress{\city{Tokyo}, \country{Japan}}}

\affil[7]{\orgdiv{Steward Observatory}, \orgname{University of Arizona}, \orgaddress{\city{Tucson}, \state{AZ}, \country{USA}}}

\affil[8]{\orgdiv{Department of Astronomy}, \orgname{University of Michigan}, \orgaddress{\city{Ann Arbor}, \state{MI}, \country{USA}}}

\affil[9]{\orgdiv{Space Science Institute}, \orgname{Lawrence Livermore National Laboratory}, \orgaddress{\city{Livermore}, \state{CA}, \country{USA}}}

\affil[10]{\orgdiv{Laboratoire Lagrange}, \orgname{Observatoire de la Côte d’Azur, Université Côte d’Azur}, \orgaddress{\city{Nice}, \country{France}}}

\affil[11]{\orgdiv{Universit\'e Paris-Saclay}, \orgname{Universit\'e Paris\textbf{} Cit\'e, CEA, CNRS, AIM}, \orgaddress{\city{F-91191 Gif-sur-Yvette}, \country{France}}}

 \abstract{

Interactions between exoplanetary atmospheres and internal properties have long been hypothesized to be drivers of the inflation mechanisms of gaseous planets and apparent atmospheric chemical disequilibrium conditions\cite{Fortney2020}. However, transmission spectra of exoplanets has been limited in its ability to observational confirm these theories due to the limited wavelength coverage of HST and inferences of single molecules, mostly H$_2$O (ref.\cite{kreidberg2018}).  In this work, we present the panchromatic transmission spectrum of the approximately 750 K, low-density, Neptune-sized exoplanet WASP-107b using a combination of HST WFC3, JWST NIRCam and MIRI. From this spectrum, we detect spectroscopic features due to H$_2$O (21$\sigma$), CH$_4$ (5$\sigma$), CO (7$\sigma$), CO$_2$ (29$\sigma$), SO$_2$ (9$\sigma$), and NH$_3$ (6$\sigma$).  The presence of these molecules enable constraints on the atmospheric metal enrichment (M/H is 10--18$\times$ Solar\cite{Lodders2009}), vertical mixing strength (log$_{10}$K$_{zz}$=8.4--9.0 cm$^2$s$^{-1}$), and internal temperature ($>$345 K). The high internal temperature is suggestive of tidally-driven inflation\cite{Millholland20} acting upon a Neptune-like internal structure, which can naturally explain the planet's large radius and low density. These findings suggest that eccentricity driven tidal heating is a critical process governing atmospheric chemistry and interior structure inferences for a majority of the cool ($<$1,000K) super-Earth-to-Saturn mass exoplanet population. }

\maketitle

The mass of WASP-107b is similar to Neptune (1.78 M$_{N}$), but its extreme low density is suggestive of a Hydrogen/Helium (H/He) envelope-to-core-mass ratio ($>$85\%; ref.\cite{Piaulet21}) more like Jupiter/Saturn (around 90\%) than like Neptune/Uranus (5-15$\%$; ref.\cite{Guillot+23_PP7}). This high of an envelope mass fraction for such a low mass planet presents challenges to the standard core-accretion paradigm of planet formation\cite{Pollack1996}---it is unclear how such a low mass planetary core (inferred to be $<$5\,M$_{\oplus}$; ref.\cite{Piaulet21}) could accrete such a massive gaseous envelope, but then stop short of fully growing into a `Jupiter'\cite{Piaulet21}. An alternative hypothesis\cite{Millholland20} is that tidal heating could inflate the planetary envelope, reducing the need for such a envelope-to-core-mass ratio.  

In addition to mass and radius constraints, measurements of the atmospheric abundances of molecules containing carbon (C), oxygen (O), nitrogen (N), and sulphur (S) can be used to jointly constrain both scenarios. The abundances of molecules in the atmosphere will be set by the intrinsic elemental inventory and the chemical processes primarily driven to disequilibrium induced by transport and photochemistry\cite{Madhusudhan2014_review, Moses2013}. The former constrains the partitioning of material between the envelope and core, while the latter is sensitive to the amount of tidal heating, altering the deep atmosphere temperature and consequently, the molecular abundances at their quenched values\cite{Fortney2020}. One of the key challenges in interpreting exoplanet atmosphere compositions is disentangling the intrinsic elemental inventory from different chemical processes, given the presence (or lack-there-of) of molecular spectral features.

Initial reconnaissance spectroscopy with the HST Wide Field Camera-3 (WFC3, 0.8--1.6\,$\mu$m) presented\cite{kreidberg2018,Spake2018} a muted water vapor absorption feature (1.4\,$\mu$m) and a surprising lack of methane (CH$_4$) absorption (1.6\,$\mu$m) --- expected to be in abundance under solar elemental abundances and chemical equilibrium at temperatures below approximately 1,000\,K (refs.\cite{Moses2013, bell2023_methane}). The presence of water but lack of methane could indicate either a low carbon-to-oxygen ratio (C/O) envelope or else be due to the quenching of methane at deeper, hotter layers\cite{Fortney2020,kreidberg2018}. The constraints on the water abundance were not precise enough to determine the bulk envelope metal enrichment. 

\begin{figure*}
    \centering
    \includegraphics[width=0.75\linewidth]{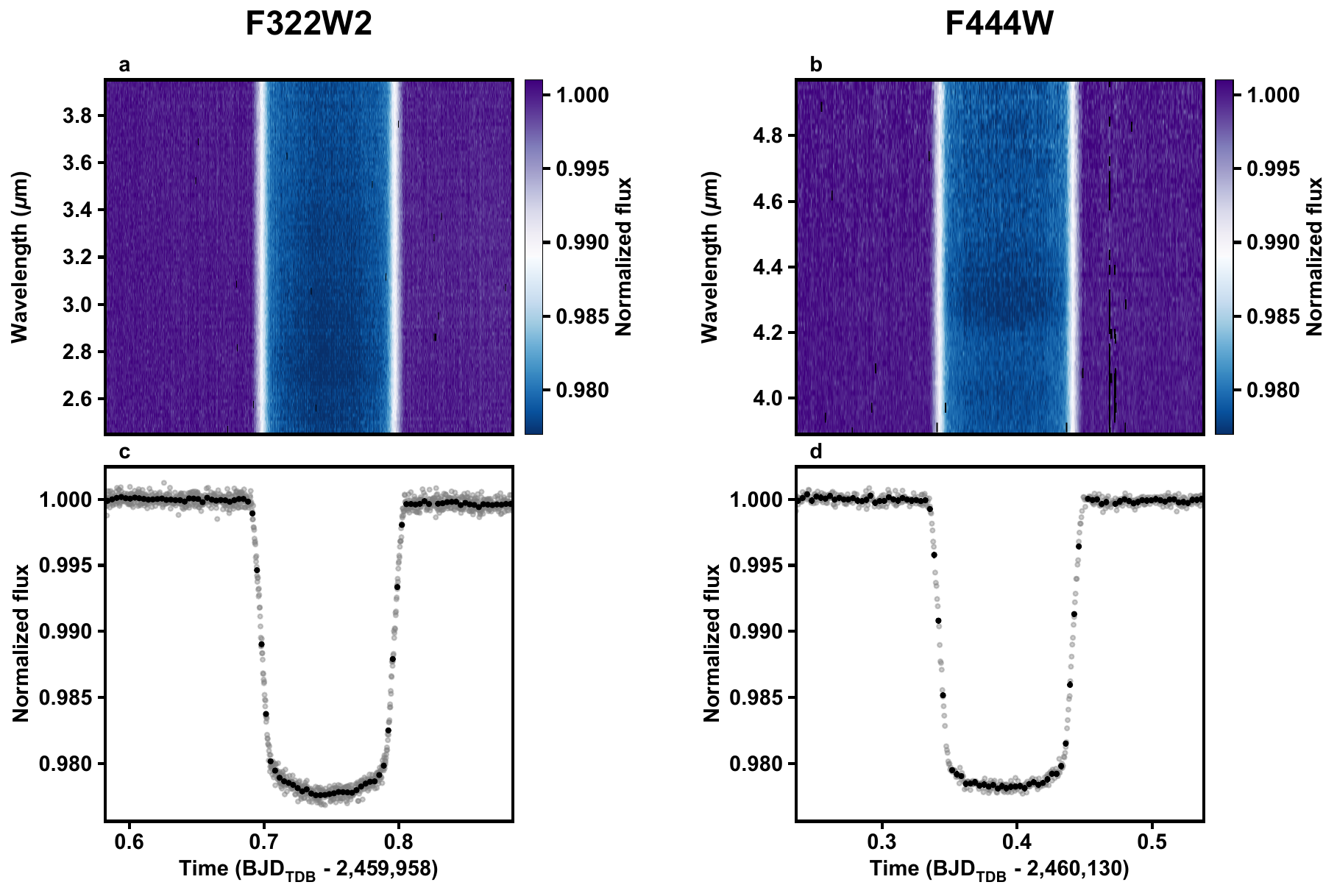}
    \caption{\textbf{Spectroscopic and broadband NIRCam lightcurves of the transit of WASP-107b.} The raw spectroscopic F322W2 and F444W transit lightcurves are shown in panels \textbf{a} and \textbf{b} after spectral binning (0.015\,$\mu$m bins) but without any temporal binning. Masked values (for example, cosmic rays) have been colored black. Even from these raw data, it is possible to visually identify the increased transit depth $\lesssim$3\,$\mu$m primarily caused by H$_2$O, from 3.9--4.1\,$\mu$m caused by SO$_2$, and from 4.3--4.5\,$\mu$m caused by CO$_2$. The F322W2 broadband (2.45--3.95\,$\mu$m) and F444W broadband (3.89--4.97\,$\mu$m) transit lightcurves are shown panels \textbf{c} and \textbf{d} with gray points without error bars. Black points with 1$\sigma$ error bars show temporally binned data with a cadence of 5 minutes; note that the error bars are typically smaller than the point size. BJD$_{\rm TDB}$ is the date in the Barycentric Julian Date in the Barycentric Dynamical Time system.}
    \label{fig:lightcurvesMain}
\end{figure*}

The broad wavelength coverage (0.6--28\,$\mu$m) offered by JWST provides access to multiple molecular bands of the major C, O, N, and S species, critical to enabling the precise atmospheric abundance constraints necessary for breaking the aforementioned degeneracies. As part of the MANATEE NIRCam+MIRI GTO program (JWST-GTO-1185; ref.\cite{Schlawin2018}), we collected two new transit observations of WASP-107b using JWST NIRCam's F322W2 (2.4--4.0\,$\mu$m) and F444W (3.9--5.0\,$\mu$m) filters and grism\cite{Greene2017} on 14 January 2023 and 4 July 2023, respectively. We analyzed the JWST/NIRCam observations with three separate data analysis pipelines (\texttt{Eureka!}\cite{bell2022}, \texttt{Pegasus} (Beatty et al., in prep.), and \texttt{tshirt}\cite{tshirt:2022}) which all agree well within error (Fig.~\ref{fig:lightcurvesMain} and Extended Data Fig.~\ref{fig:nircamSpectraMethods}); we ultimately chose to adopt \texttt{Eureka!}'s analysis for our fiducial NIRCam spectrum when performing atmospheric modelling. In addition, we incorporate recently published JWST/MIRI LRS observations\cite{dyrek2023} as well as the previously published HST/WFC3 G102 and G141 observations\cite{kreidberg2018,Spake2018}. To ensure a uniform set of orbital and limb-darkening parameters between the many different instruments, we re-analyzed the HST/WFC3 observations using \texttt{Pegasus} and the JWST/MIRI LRS observations using \texttt{Eureka!}. We find that these re-analyzed WFC3 and MIRI/LRS spectra are consistent with the literature spectra aside from a constant offset caused by the improved orbital solution. All together, these data give us a panchromatic dataset with observations spanning 0.8--12.2\,$\mu$m, with continuous coverage between 2.45\,$\mu$m and 12.2\,$\mu$m.

\begin{figure}
    \centering
    \includegraphics[width=\linewidth]{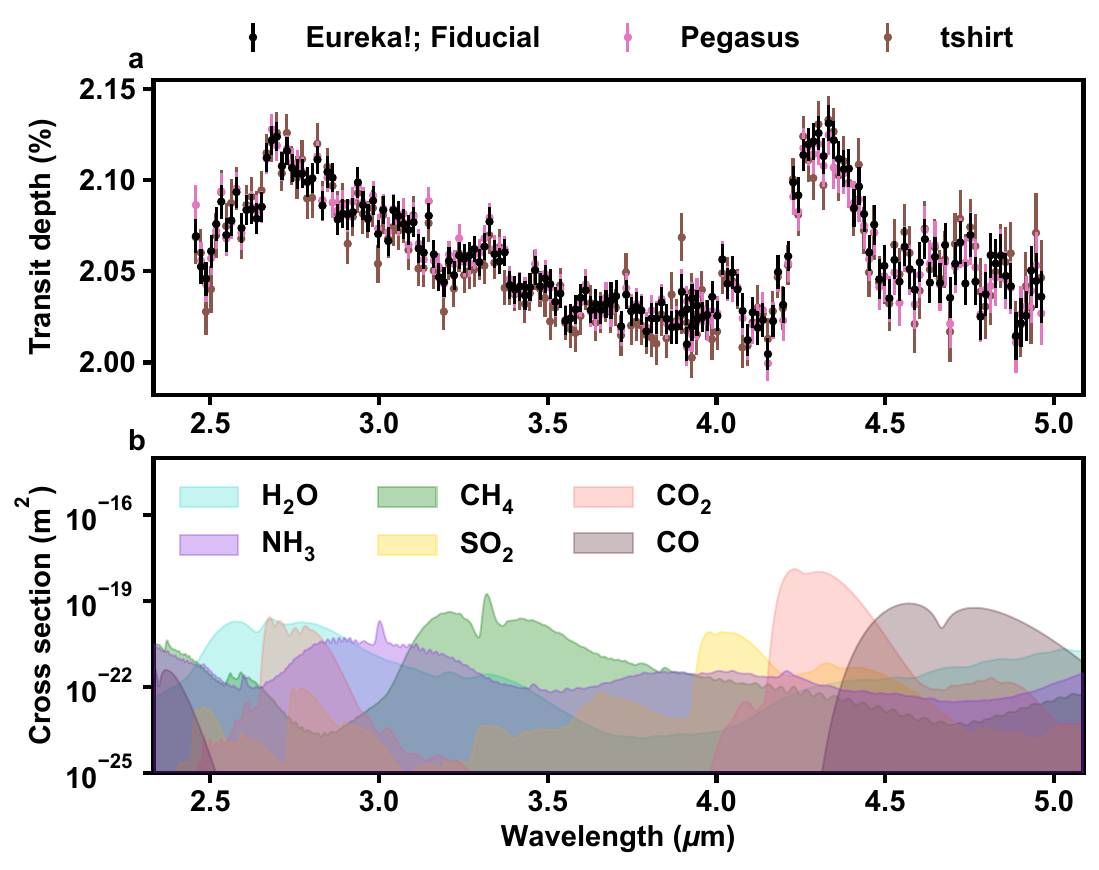}
    \caption{\textbf{Independent reductions of the transmission spectrum of WASP-107b.} \textbf{a,} The NIRCam F322W2 and F444W transmission spectra as reduced by the \texttt{Eureka!}, \texttt{Pegasus}, and \texttt{tshirt} pipelines are shown in different colors at a constant wavelength binning of $\Delta \lambda = 0.015$\,$\mu$m and with 1$\sigma$ error bars. All three analyses show clear agreement on the overall shape of the spectrum, with features from H$_2$O (${\sim}$2.5--3.5\,$\mu$m), CH$_4$ (${\sim}$3.2--3.8\,$\mu$m), SO$_2$ (${\sim}$3.9--4.1\,$\mu$m), and CO$_2$ (${\sim}$4.2--4.6\,$\mu$m) all clearly visible by-eye in each of the reductions. \textbf{b,} The absorption cross sections of the six detected species, many of which are visually identifiable in panel \textbf{a}.}\label{fig:nircamSpectraMain}
\end{figure}

\begin{figure*}
    \centering
    \includegraphics[width=0.75\linewidth]{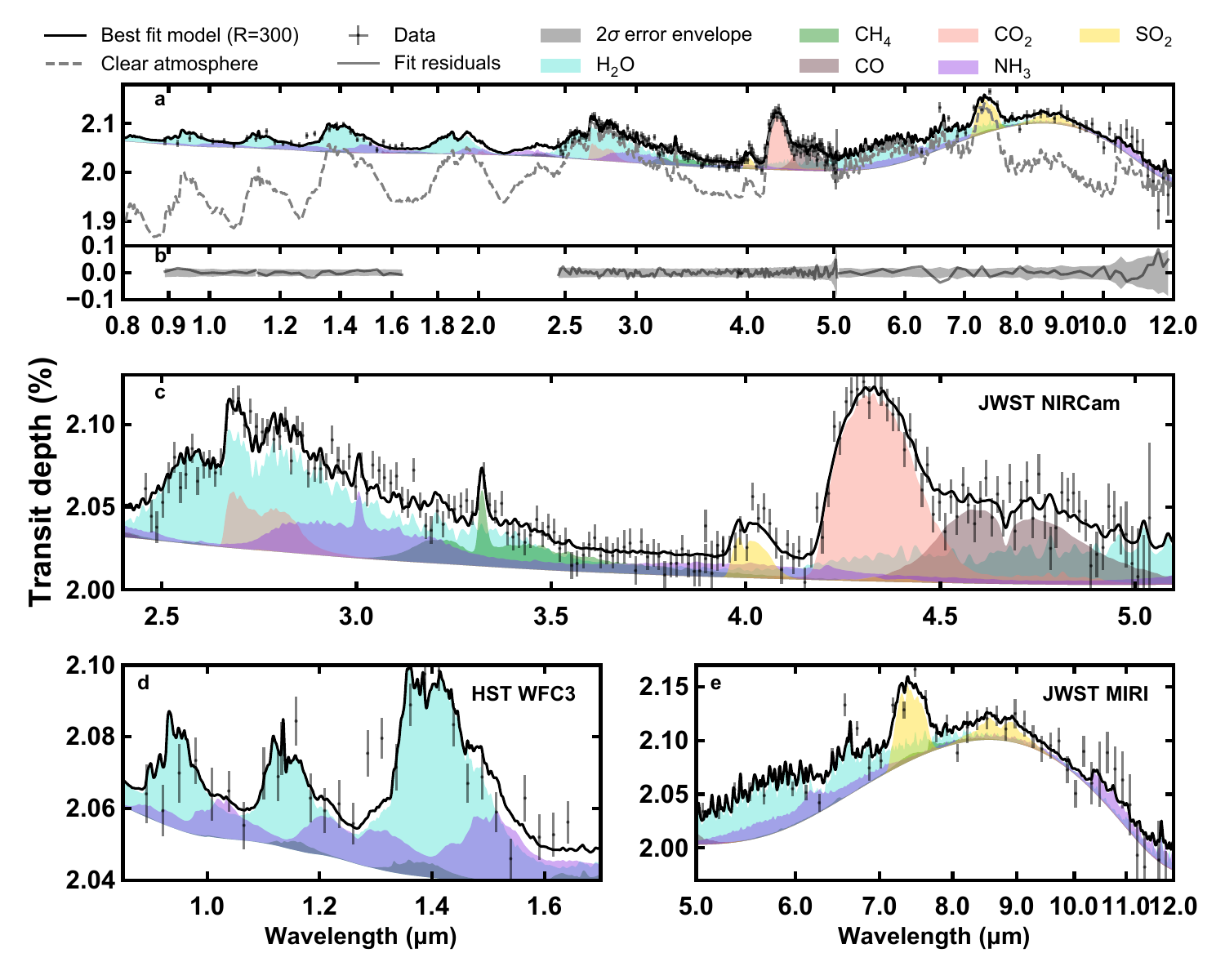}
    \caption{\textbf{Interpretation of WASP-107b's transmission spectrum.} The observed transmission spectrum of the planet with 1$\sigma$ error bars is compared to the best-fit one-dimensional 
    radiative-convective-photochemical equilibrium models (1D-RCPE model, $\chi^2/N_{\rm data}$=1.4) shown at a spectral resolution of R=300. The colored shaded regions show the contributions from individual gases to the best fit model. The gray dashed line shows the clear atmosphere component of the model, that is, the gas contributions without the presence of clouds or hazes. Panel \textbf{a} shows the broad wavelength spectrum (0.8--12\,$\mu$m) of the planet. Panel \textbf{b} shows the fit residuals with the corresponding 2$\sigma$ data error envelope. Panel \textbf{c} shows the NIRCam observations, while \textbf{d} shows HST WFC3 spectra and \textbf{e} shows JWST MIRI observations. The HST/WFC3 data are binned at a constant $\Delta \lambda = 0.025$\,$\mu$m, the NIRCam data are binned at $\Delta \lambda = 0.015$\,$\mu$m, and the MIRI data are binned at $\Delta \lambda = 0.15025$\,$\mu$m.} \label{fig:spectrum_and_models}
\end{figure*}

Among the chemical species expected in exoplanet atmospheres\cite{Moses2013, Madhusudhan2019} (see Methods), the final NIRCam spectrum shown in Figure \ref{fig:nircamSpectraMain} shows prominent absorption features due to H$_2$O (2.5--3.2\,$\mu$m, detected at 21$\sigma$), CO$_2$ (2.66--2.86\,$\mu$m and 4.2--4.5\,$\mu$m, detected at 29$\sigma$), CO (4.5--4.9\,$\mu$m, detected at 7$\sigma$), and SO$_2$ (3.94--4.1\,$\mu$m, detected at 9$\sigma$), with weaker features due to CH$_4$ (3.2--3.5\,$\mu$m, detected at 5$\sigma$) and NH$_3$ (2.9--3.1\,$\mu$m, detected at 6$\sigma$), with reported detections from the 1-dimensional radiative-convective-photochemical equilibrium models (1D-RCPE) described below. These features are complemented by two additional H$_2$O bands in HST/WFC3 (around 1.13\,$\mu$m and 1.4\,$\mu$m) and another in MIRI (between 5--7\,$\mu$m), along with another strong SO$_2$ feature (7.1--7.7\,$\mu$m) and a weak NH$_3$ feature (10.3--11\,$\mu$m) in MIRI, shown in Figure \ref{fig:spectrum_and_models}. The panchromatic spectrum exhibits a slight downward slope from blue-to-red, indicative of aerosol scattering\cite{Lecavelier2008a,Sing2016, OhnoKawashima20_slope}, and a strong concavity across MIRI, previously attributed to silicate cloud particulate resonance features\cite{dyrek2023}.  The presence of both CO$_2$ and SO$_2$ features is indicative of envelope metal enrichment above solar and photochemistry\cite{Tsai2022}, while the relatively weak CH$_4$ feature is suggestive of CH$_4$ depletion\cite{kreidberg2018, Spake2018, dyrek2023}.

To rigorously infer the atmospheric composition and internal temperature from the observed transmission spectrum, we employ Bayesian inference with the transmission spectra generated from a suite of 1-dimensional radiative-convective-photochemical equilibrium models (1D-RCPE)\cite{bell2023_methane} (see Methods). This self-consistent method properly captures the degeneracies between the intrinsic atmospheric composition and chemical processes. The final result from this process are samples from a posterior-probability distribution that represents the constraints on T$_{\rm irr}$, T$_{\rm int}$, [M/H], C/O, and cloud properties, and an estimate of the Bayesian model evidence, which we use to draw conclusions about WASP-107b's atmosphere. Figure \ref{fig:spectrum_and_models} shows that this model setup adequately explains the observed panchromatic spectrum, capturing nearly all of the salient features. From this process we find a metallicity of $10\text{--}18\times$ solar\cite{Lodders2009} (at 68$\%$ confidence, median of $12.3\times$ solar) and a sub-solar carbon-to-oxygen ratio (C/O$=0.33^{+0.06}_{-0.05}$). Furthermore, the spectrum also requires a high internal temperature (T$_{\text{int}}>345$\,K at $99.7\%$ confidence, while a value of $<100$~K would be expected\cite{Millholland20} given the planet's low mass and the star's $\sim$3 Gyr age\cite{Piaulet21}), and an atmosphere with a strong enough vertical mixing (that is, eddy diffusion, $\log(K_{zz})=8.6^{+0.4}_{-0.2}$) to quench methane along the deeper (0.25--0.65\,bar) and hotter (around 1,100--1,300\,K) parts of the atmosphere, confirming previous suggestions\cite{kreidberg2018, dyrek2023}.

\begin{figure}
    \centering
    \includegraphics[width=\linewidth]{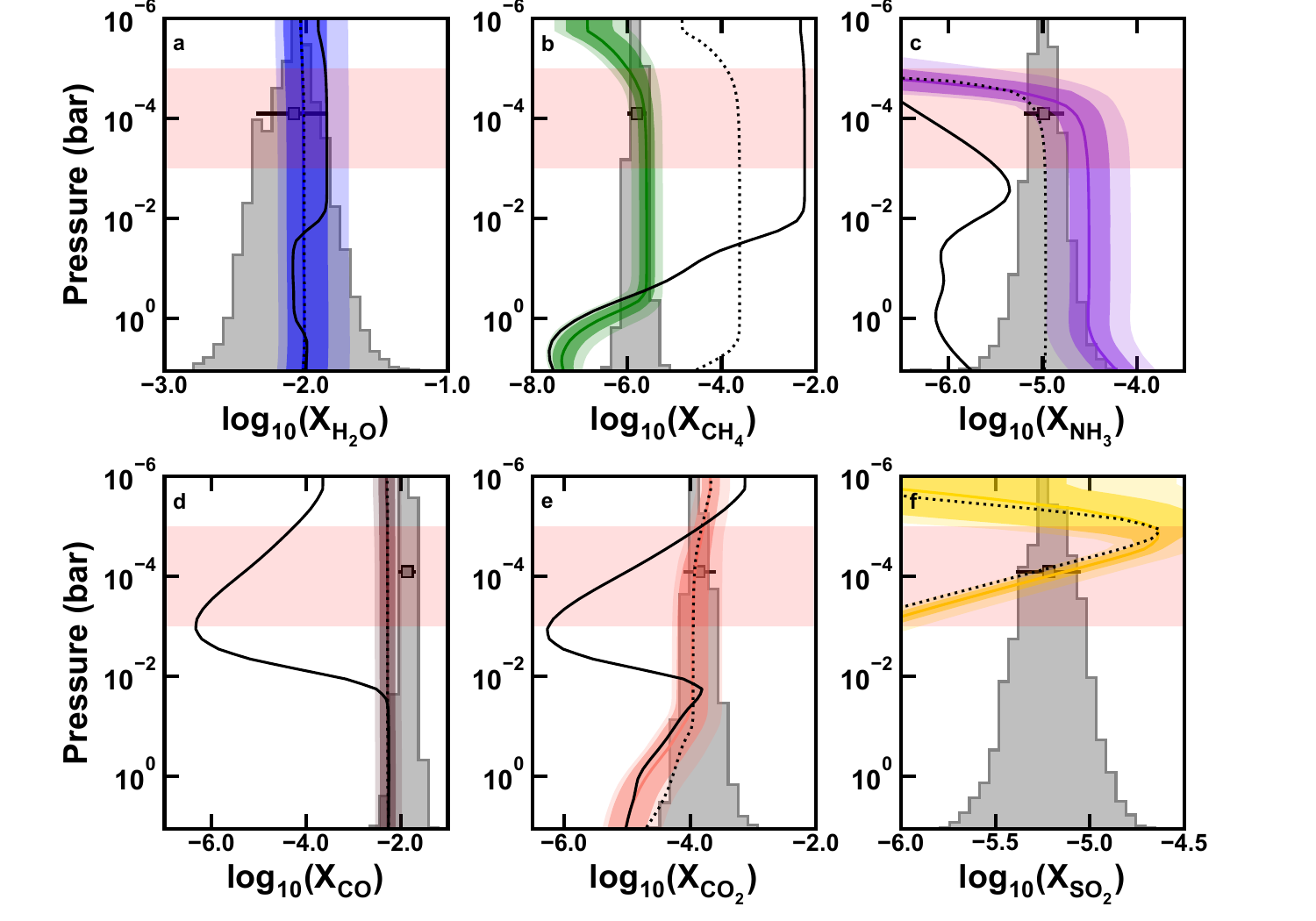}
    \caption{\textbf{Inferred molecular volume mixing ratios from WASP-107b's transmission spectrum.} The retrieved volume mixing rations (that is, abundances) for each detected gas (panels \textbf{a}--\textbf{f}) are shown from the free-retrieval (gray posterior distributions, with median and 68\% confidence error bars) and from the 1D-RCPE grid inference (median and 68\% confidence profiles shown as colored pressure-dependent abundance profiles). The black solid lines show expectations for an atmosphere in chemical equilibrium under the inferred [M/H] and C/O. The dotted black lines show expectations for chemical disequilibrium under a cool internal temperature of 200~K.  The self-consistent abundance profiles from the 1D-RCPE retrieval for a high internal temperature are generally consistent with the retrieved abundances from the free retrieval within 1$\sigma$. The red horizontal shaded region spans pressures from 1mbar to 10$^{-5}$ bar and corresponds to the pressures probed by our observations. The colors of the abundance profiles match the contributions in Figure \ref{fig:spectrum_and_models} and the cross-sections in panel b of Figure \ref{fig:nircamSpectraMain}. }\label{fig:grid_v_free}
\end{figure}

We further validate the interpretation from the self-consistent 1D-RCPE models by performing additional Bayesian inferences using parametric atmospheric models that independently fit for the chemical abundances and vertical pressure-temperature structure of the planetary atmosphere without any assumptions of radiative-convective thermo-chemical equilibrium (called, `free-retrievals'). Using two independent inference frameworks (Aurora and CHIMERA, see Methods, detection significances from CHIMERA) we confirm the detections of CO$_2$ (27$\sigma$), H$_2$O (18$\sigma$), SO$_2$(8$\sigma$), CO (5$\sigma$), NH$_3$ (5$\sigma$), and CH$_4$ (8$\sigma$) at strong confidence in agreement with the 1D-RCPE models. We find that the detection of NH$_3$ is mostly driven by the NIRCam observations and do not depend solely on the red-edge of the MIRI observations (see Methods). The derived chemical abundances, assumed to be constant with height by these models, are also consistent with the inferred abundance profiles from the 1D-RCPE models (see Fig.~\ref{fig:grid_v_free}) and are also suggestive of a metal enriched envelope with depleted CH$_4$ abundances.

The relatively high internal temperature we infer for WASP-107b is likely caused by tidal heating in the planetary interior. Recent radial-velocity observations measured the planet to have a mildly non-zero orbital eccentricity of $e=0.06\pm0.04$ (ref.\cite{Piaulet21}). Tidal heating relations\cite{LeconteTidalHeating} predict that at this nominal $e=0.06$ the internal temperature of WASP-107b will be T$_{\text{int}}\approx350$\,K for a Neptune-like tidal quality factor of $Q = 10^4$. The expected internal temperature from tidal heating drops to T$_{\text{int}}\approx200$\,K to T$_{\text{int}}\approx110$\,K for Jupiter-like tidal quality factors of $Q = 10^5$ to $Q=10^6$. Additional heating of the deep atmosphere may be achieved by vertical and horizontal mixing\cite{Showman2002, Sainsbury-Martinez2019, Sarkis2021, Schneider2022}. If we assume that tidal effects are solely responsible for heating WASP-107b's atmosphere, then our retrieved 3$\sigma$ lower limit of T$_{\text{int}}>345$\,K implies a corresponding 3$\sigma$ upper limit on the planet's tidal quality factor of $Q < 10^{3.8}$. If WASP-107b has a tidal quality factor significantly lower than Neptune ($10^{3.9} \lesssim Q_N \lesssim 10^{4.5}$),\cite{NeptuneTidalQ} this would be consistent with the inference of a large core for the planet, since typical core material for ice giants is expected to have $Q \approx 10^2$, while typical gas envelopes have $Q \approx 10^5$ \cite{millholland19}

The large value of T$_{\rm int}$ implies a much hotter, lower-density H/He envelope that has been previously appreciated.  Using state-of-the-art structure models\cite{Tang2023}, along with T$_{\rm int}= 350$~K (the retrieved 3$\sigma$ lower limit), we find that the planet's radius can be explained by a model that has $>$22~$M_{\oplus}$ of rock/iron in its interior, here modeled as a distinct core.  This ratio of solids to H/He is similar to that of the bulk composition of Uranus and Neptune in our solar system\cite{Helled2020}.  Our revised view of the planet is quite different than previous work\cite{Piaulet21,Anderson2017}, as without tidal heating the planet's low bulk density can only be explained by a structure that is mostly H/He with a very small core ($<5M_{\oplus}$, ref\cite{Piaulet21}), perhaps at odds with core-accretion theory. 

The inferred atmospheric metallicity of WASP-107b is lower than expectations from the Solar System metal enrichment trend\cite{Welbanks2019b} as determined by the CH$_4$ abundances\cite{Atreya2018}, which predicts an enhancement of about 32$\times$ solar (see also ref.\cite{Kreidberg2014b}). Our results confirm that different elements in a planetary atmosphere can be differently enhanced, as previously suggested by HST observations\cite{Welbanks2019b}. Furthermore, our results demonstrate that individual molecules (for example, H$_2$O) may not be good bulk metallicity tracers. Instead, JWST broadband spectra as presented here, presents a key opportunity to derive estimates of the bulk atmospheric metal enrichment informed by the measured abundances of several gases and by self-consistently considering the impact of disequilibrium processes arising from interactions between the interior and atmosphere of the planet. 

The understanding of vertical mixing in the atmospheres of giant planets and brown dwarfs remains a significant challenge in atmospheric physics\cite{Fortney2020,Miles2020}. The inferred vertical mixing strength ($K_{zz}$) can be high ($K_{zz}{\sim}10^8\text{--}10^{11}$ cm$^2$\,s$^{-1}$, depending on T$_{\rm int}$) in convective zones where mixing is driven by convective overturn and mixing length theory can be used, or much more uncertain in radiative zones where the driving mechanism remains unclear\cite{Menou2019, Komacek2019}. Recent studies of mixing in brown dwarf atmospheres, at temperatures similar to WASP-107b\cite{Miles2020, Mukherjee2024}, infer low $K_{zz}$ values of 10$^2$--10$^5$ cm$^2$\,s$^{-1}$ due to sluggish mixing in deep atmospheric radiative zones where chemical abundances are quenched. Studies of this same phenomenon in Jupiter have long suggested a $K_{zz}$ value of ${\sim}10^8\text{--}10^9$ cm$^2$\,s$^{-1}$, although recent work\cite{Cavalie2023} points to a lower value ($K_{zz}\sim10^6$ cm$^2$\,s$^{-1}$) as the best explanation for Jupiter's CO and H$_2$O abundances, again perhaps due to a deep radiative zone. Overall, the high $K_{zz}$ values implied for WASP-107b are striking in comparison to other objects. This suggests the quenching of chemical abundances in an atmospheric convective zone, with a high internal flux and shallow radiative-convective boundary.  This can only be achieved if significant additional internal energy, as from tidal heating, keeps the interior and deep atmosphere much warmer than a standard cooling model would suggest.

The detection of the major C, O, N, and S reservoirs in the spectrum of WASP-107b resulting from the broad wavelength coverage demonstrates the unparalleled capabilities of JWST for the detailed atmospheric characterization of exoplanet atmospheres. The simultaneous constraints on multiple chemical species enables unique solutions for the bulk atmospheric metal enrichment, strength of vertical mixing, and provides strong evidence for a high internal temperature most likely arising from tidal heating due to a small, yet significant, orbital eccentricity. The combination of constraints on atmospheric metallicity and the internal heat flux together provide novel constraints on the relative gas-to-core mass fraction, naturally explaining the low planetary density. These initial constraints on the elemental ratios of C, O, S, and N have the potential to inform the planetary accretion history\cite{Crossfield2023}.  

The detection and constraint on the abundance of CH$_4$ in this warm Saturn adds to the growing number of inferences of this sought-after carbon-bearing species in transiting exoplanets\cite{bell2023_methane, Madhusudhan2023}. Furthermore, the low abundance of methane relative to thermochemical expectations confirms its sensitivity to mixing processes and the internal temperature/heat flux\cite{Fortney2020}, providing a critical atmospheric diagnostic of interior processes and structure.  We anticipate that most common type of planets across the galaxy will have elevated internal temperatures arising from tidal heating, and thus should have relative depletions of methane. Of the 262 known cool ($<$1,000\,K) `Neptune-like'\cite{ChenKipping2017} planets (with well determined mass and radii), approximately 2/3 have a reported non-zero eccentricity\cite{Southworth2011}. As with WASP-107b, deciphering the intrinsic nature of these worlds will be critically dependent upon the constraints on the total elemental inventory, the strength of vertical mixing, and the internal temperature. Future studies from the MANATEE program will explore the transmission and emission spectrum of warm exoplanets to further disentangle the effects of disequilibrium chemistry due to interior-atmosphere interactions, photochemistry, and directly link this population to the solar system gas giants and their theorized formation pathways.

\bibliographystyle{sn-standardnature}

\clearpage
\backmatter

\section*{Methods} \label{sec:methods}
\renewcommand{\figurename}{Extended Data Fig.}
\renewcommand{\tablename}{Extended Data Table}
\renewcommand{\theHfigure}{Extended Data Fig.~\arabic{figure}}
\renewcommand{\theHtable}{Extended Data Table \arabic{table}}
\setcounter{figure}{0}
\setcounter{table}{0}

\subsection*{NIRCam Reduction} \label{sec:nircamReduction}

Both of our new NIRCam observations used the BRIGHT2 readout pattern and the SUBGRISM256 subarray, with the F322W2 observations using 7 groups per integration and 1,293 integrations while the F444W observations used 15 groups per integration and 625 integrations. In the subsections below, we will describe the three independent pipelines we used to reduce these NIRCam data.

\paragraph{Eureka!}\label{sec:eurekaReduction}

Our \texttt{Eureka!} reduction used version 0.9 of the \texttt{Eureka!} pipeline\cite{bell2022}, CRDS version 11.17.0 and context `1093' (for F322W2) or `1097' (for F444W), and \texttt{jwst} package version 1.10.2 (ref.\citeApp{jwst_v1.10.2}). Our reduction methods closely follow those used in previous \texttt{Eureka!} NIRCam spectroscopy analyses\cite{bell2023_methane}$^,$\citeApp{ahrer2022nircam}. The \texttt{Eureka!} Control Files and \texttt{Eureka!} Parameter Files we used are available for download from Zenodo (\url{https://doi.org/10.5281/zenodo.10780448}; ref.\citeApp{welbanks2024_wasp107b_zenodo}), and the important parameters are summarized below.

We ran \texttt{Eureka!}'s Stages 1 and 2 on both the F322W2 and F444W observations using the default \texttt{jwst} processing steps with the exception of increasing the jump rejection threshold in Stage 1 to 6$\sigma$ from the default of 4$\sigma$ (to avoid excessive false positives) and turning off the photom step in Stage 2 (as flux-calibrated data is not desired for time-series observations). In \texttt{Eureka!}'s Stage 3, we first did two iterations of 5$\sigma$ clipping along the time axis on the background pixels, and we also masked pixels marked as `DO\textunderscore NOT\textunderscore USE' in the data quality (DQ) array. Next, we corrected for the curvature of the trace by rolling the pixels along the spatial axis in integer pixels (to avoid the need to interpolate or sub-sample the pixels) until the whole trace was approximately centered on the subarray. We then subtracted the background flux from each column in each integration by fitting a linear slope to the column after masking pixels within 13 pixels of the center of the star's trace and masking any 7$\sigma$ outliers in the column. The spatial position and PSF-width of the star were then recorded for each integration by first summing along the spectral axis and then fitting a Gaussian profile; these parameters were not used in the reduction step but were instead recorded as potential covariates when fitting the lightcurves in \texttt{Eureka!}'s Stage 5. We then performed optimal spectral extraction\citeApp{horne1986optspec} using only the pixels within 5 pixels of the center of the spectral trace and using a cleaned version of the median integration to compute our spatial profile; to compute this profile, we clipped 5$\sigma$ outliers along the time axis and then smoothed along the spectral direction using a boxcar filter with a width of 13 pixels. Pixels which differed by more than 10$\sigma$ compared to the spatial profile were clipped when performing optimal extraction to remove cosmic rays missed during the Stage 1 processing. In Stage 4, we first removed wavelengths with excessively high noise (likely due to unmasked bad pixels); we masked 15 wavelengths in the F322W2 data and one in the F444W data. We then spectrally binned the data into 0.015\,$\mu$m wide bins for both NIRCam filters (100 bins spanning 2.45--3.95\,$\mu$m for F322W2 and 72 bins spanning 3.89--4.97\,$\mu$m for F444W). Finally, to remove any remaining cosmic ray effects, we clipped any 4$\sigma$ outliers in the spectrally binned lightcurves compared to a cleaned version computed using a boxcar filter with a width of 20 integrations which was narrow enough to not clip the ingress or egress of the transit.

\paragraph{Pegasus}\label{sec:pegasusReduction}

We also reduced and extracted the transmission spectrum of WASP-107b using the \texttt{Pegasus} pipeline (\url{https://github.com/TGBeatty/PegasusProject}). We will describe \texttt{Pegasus} in more detail in a forthcoming paper (Beatty et al., in prep.), and we briefly summarize it here. Our \texttt{Pegasus} reduction started from the \textsc{rateints} files from the \texttt{jwst} pipeline v1.10.2, using CRDS version 11.17.0 and context `1093'.

We first performed a background subtraction step for each \textsc{rateints} file by fitting a two-dimensional second-order spline to each integration using the entire 256$\times$2,048 \textsc{rateints} images. We masked out image rows 5 to 75 to not self-subtract light from the WASP-107 system, and then we fit individual background splines to each of the four amplifier regions in the images. We then extrapolated the combined background spline over the masked portions near the star to perform the background subtraction. Visual inspection of the \textsc{rateints} images also showed that in roughly 5\% of the integrations the reference pixel correction failed for at least one of the amplifier regions, so after the spline fitting and subtraction we re-ran the reference pixel correction using \texttt{hxrg-ref-pixel} (\url{https://github.com/JarronL/hxrg_ref_pixels}). This appeared to correct the issue.

We then performed a spectrophotometric extraction on the background-subtracted images using optimal extraction techniques\citeApp{horne1986optspec}. We iteratively constructed a smoothed spatial profile for the F322W2 data using pixel columns from columns 22 to 38 (inclusive) and pixel rows from rows 5 to 1,650 (inclusive), and we iteratively estimated a variance matrix for this same region starting from the pixel uncertainties and read-noise values following established techniques\citeApp{horne1986optspec}. We then fit a 4th-order polynomial spectral trace to each F322W2 integration and extracted fluxes in each wavelength bin using an extraction aperture with a half-height of seven pixels centered on the estimated trace in each detector column. In doing so, we accounted for partial pixel effects in both the spectral and spatial directions.

\paragraph{tshirt}\label{sec:tshirtReduction}
We reduced the data using the Time Series Helper \& Integration Reduction Tool \texttt{tshirt}\cite{tshirt:2022} (\url{https://github.com/eas342/tshirt}), which uses modified steps of the \texttt{jwst} pipeline to calculate rate images and performs a custom co-variance weighted optimal extraction\citeApp{schlawin2020jwstNoiseFloorI}.
We followed a very similar procedure as previous NIRCam reductions\cite{bell2023_methane}$^,$\citeApp{ahrer2022nircam}.
We begin by running the initial data quality, saturation and superbias subtraction steps with default parameters and reference of the \texttt{jwst} package\citeApp{jwst_v1.10.2} version 1.10.2 and CRDS version 11.16.22.
For Observation 8 (using the F322W2 filter), CRDS context \texttt{jwst\_1137.pmap}, which we re-ran at a later time.
For Observation 9 (using the F444W filter), we context (\texttt{jwst\_1093.pmap}).
Despite the CRDS context numbers, we verified that all reference files were the same between the two filters for Stage 1 processing.
We next used a the row-by-row odd/even by amplifier (ROEBA) step\citeApp{schlawin2023defocused} to reduce the 1/$f$ noise by subtracting sky pixels across the spectrum (for the long wavelength grism) and defocused PSF.
We continue the rest of the processing of the \texttt{calwebb\_detector1} pipeline with only modification of setting the jump step's rejection threshold to 6$\sigma$.
We then use the imaging flat field \texttt{jwst\_nircam\_flat\_0313.fits} and divide the \texttt{nrcalong} images by this for flat field correction.

For the spectroscopy, the background subtraction was achieved with a column-by-column fit to the regions from 10 to 30 pixels on either side of the source.
A robust (sigma-clipped) line was fit to each column of each integration and the line was subtracted from the column.
We use co-variance weighted extraction\citeApp{schlawin2020jwstNoiseFloorI} to sum the pixels in the cross-dispersion direction with an aperture full width of 10 pixels.
For the covariance matrix, we assume a constant correlation of 0.08 across pixels.
The median measured standard deviation across integration 10 and integration 100 for the F322W2 data with 15 pixel-wide wavelength bins is 1,070 ppm compared to a theoretical error (from photon and read noise) of 990 ppm.
The median measured standard deviation across integration 10 and integration 100 for the F444W data with 15 pixel-wide wavelength bins is 985 ppm compared to a theoretical error (from photon and read noise) of 971 ppm.
We note that the F444W has better 1/$f$ subtraction from sky pixels on 2 amplifiers whereas the F322W2 only has sky pixels available one one amplifier. The noise is elevated from theoretical expectations because there are still 1/$f$ correlations in the fast-read direction of the detector.

\subsection*{HST/WFC3 Re-reduction}

HST previously observed transits of WASP-107b with Wide-Field Camera 3 (WFC3) using the G141 and G102 dispersers. The results of these observations have been reported previously\cite{kreidberg2018,Spake2018}. For consistency with the rest of our data analysis, we re-reduced and re-analyzed these data.

The G141 data cover a single transit of WASP-107b on UT 6 June 2017 using five orbits of HST time as a part of HST program GO-14915 (P.I. L. Kreidberg). At the start of each orbit, HST took a single F130N direct image of WASP-107 to establish a wavelength reference, and then the observatory switched to spatially scanned spectroscopic observations using the G141 grism. The full details of the observational setup are described in the original report on the results of this observation\cite{kreidberg2018}.

The shorter wavelength G102 data also cover a single transit of WASP-107b using five orbits of HST time on UT 31 May 2017 as a part of HST program GO-14916 (P.I. J. Spake). Each orbit of the G102 transit observation began with an F126N direct image to create a wavelength reference, and then the G102 grism was used to collect spatially scanned spectroscopy of the WASP-107 system. The complete description of these observations is reported in the original paper describing these data\cite{Spake2018}.

We reduced both transit observations using a custom data reduction pipeline that has been described elsewhere\citeApp{beatty2017kepler13} and which we briefly summarize here. For each spatially scanned exposure, we began by extracting 1D spectra for each of the individual up-the-ramp samples, accounting for both the residual background light and the slight angle of the scan pattern on the detector. We then summed these individual spectra to create the 1D spectrum from that exposure. We used the wavelength solutions generated from the direct images taken at the start of each orbit to extract spectroscopic lighturves in wavelength bins matching those used in the previous works on these data\cite{kreidberg2018,Spake2018}. These were nine evenly spaced bins from 0.877--1.139 $\mu$m for the G102 data and twenty evenly spaced bins from 1.121--1.629 $\mu$m for the G141 data. We note that with the updated transit ephemeris for WASP-107b from these observations, we do not see evidence for a starspot crossing in the WFC3 data as was reported previously,\cite{kreidberg2018} nor do we see starspot crossings in the NIRCam or MIRI observations.

\subsection*{MIRI Re-reduction}\label{sec:miriReduction}

We decided to independently re-reduce the MIRI/LRS spectra of ref.\cite{dyrek2023} to ensure the reduction was robust to different reduction choices and to ensure a self-consistent reduction of the NIRCam and MIRI data. Our \texttt{Eureka!} re-reduction of the MIRI/LRS observations used version 0.9 of the \texttt{Eureka!} pipeline\cite{bell2022}, CRDS version 11.17.0 and context `1097', and \texttt{jwst} package version 1.10.2 (ref.\citeApp{jwst_v1.10.2}). Our reduction method generally follows the \texttt{Eureka!} v1 method described by ref.\citeApp{bell2023_ers}, with some differences and experimental new steps. The \texttt{Eureka!} Control Files and \texttt{Eureka!} Parameter Files we used are available for download from Zenodo (\url{https://doi.org/10.5281/zenodo.10780448}; ref.\citeApp{welbanks2024_wasp107b_zenodo}), and the important parameters are summarized below.

In our Stage 1 processing, we turned on the firstframe and lastframe steps (which respectively discard the first and last frames in each integration) in the \texttt{jwst} pipeline and increased the jump rejection threshold to 7$\sigma$ from 4$\sigma$ as we found these changes resulted in reduced scatter in the residuals of our lightcurve fits. We had also considered different combinations of turning on/off the firstframe, lastframe, and RSCD correction steps but found that it was best to turn the firstframe and lastframe steps on and leave the RSCD step off for these data. We also investigated using the default CRDS linearity reference file vs the linearity reference file used by ref.\cite{dyrek2023} and found that there was no clear impact on our final transmission spectra, so we adopted the linearity reference file used by ref.\cite{dyrek2023}.

As pointed out by ref.\citeApp{bell2023_ers}, it appears that narrow wavelength bins result in excessively noisy transmission spectra and underestimated error bars for the MIRI/LRS time-series observation (TSO) commissioning target L168-9b. Ref.\citeApp{bell2023_ers} hypothesized that this wavelength-correlated noise could be the result of 390\,Hz periodic noise observed in some MIRI subarrays (ref.\citeApp{Bouwman2023MiriTsoCommissioning}; private comm., Michael Ressler). This noise appears as structured noise with a period of around 9 rows within an individual group and is likely caused by MIRI's electronics. To investigate the potential impacts of this structured noise on MIRI time-series observations, we developed a experimental 390\,Hz noise removal method for the \texttt{Eureka!} pipeline. This method takes advantage of the knowledge that the noise exhibits a fixed waveform with the phase of the periodic noise remaining fixed within an integration but varying between integrations (only varying slightly and with an apparent cycle of 8 integrations); we determined the best-fit waveform using an entire segment of our science observations, and then we freely fit the phase shift of the waveform for each integration. We found that there is significant power in the 1st, 2nd, and 4th harmonics with the 1st harmonic's frequency being exactly 390.625\,Hz ($f$=1/(256$\times$10\,$\mu$s); private comm., Michael Ressler). We found that our new algorithm significantly reduced the periodic noise visible in a Lomb-Scargle periodogram\citeApp{Lomb1976,Scargle1982} at the group level. While performing this 390\,Hz noise removal step, we also performed group-level background subtraction (GLBS) for each row in each group (MIRI's dispersion direction is along a column, so rows run in the spatial direction) using the mean value of columns 11--30 and 44--63 (avoiding pixels contaminated by the star); this further removed structured noise from the Lomb-Scargle periodogram. To understand the impact on our final transmission spectra, we also performed another reduction without the GLBS step and another without either the 390\,Hz noise removal or GLBS steps.

\begin{figure*}
    \centering
    \includegraphics[width=0.85\linewidth]{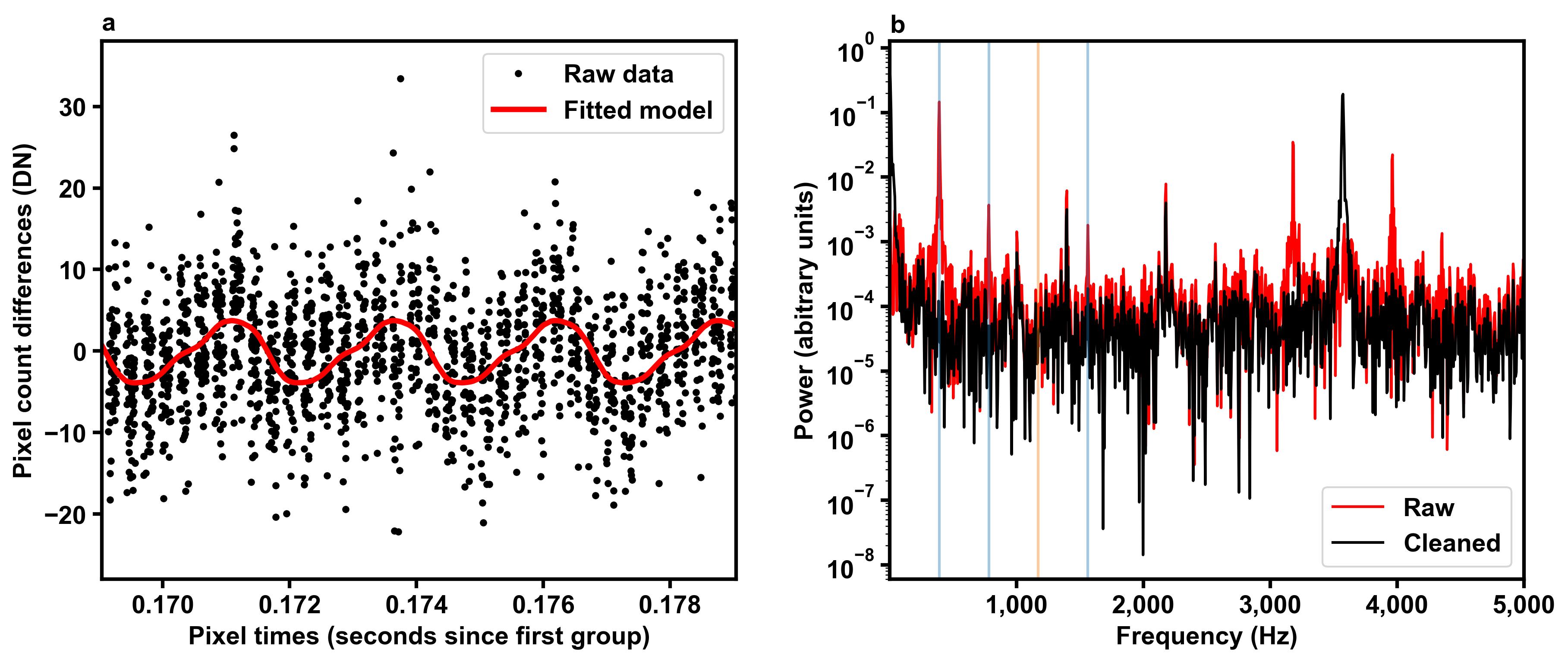}
    \caption{\textbf{MIRI/LRS's 390Hz noise signal visualized.} \textbf{a:} The background pixel values for part of the second integration of the MIRI/LRS observations of WASP-107b are shown in black points, while our fitted 390\,Hz noise signal is shown with a red line. \textbf{b:} The Lomb-Scargle periodogram of the pixels in the second group of the second integration before the 390\,Hz noise removal and GLBS steps are applied are shown in red, and the periodogram after these two steps are applied is shown in black. The three harmonics used in our cleaning procedure (1st, 2nd, and 4th) are indicated with pale blue vertical lines, while skipped 3rd harmonic is shown with a pale orange vertical line. The clear impact of the 390\,Hz noise removal and GLBS steps is seen in the eliminated spikes in the Lomb-Scargle periodogram.}
    \label{fig:390hz_signal}
\end{figure*}

In \texttt{Eureka!}'s Stage 2, we turned off the photom step as flux-calibrated spectra are not desired for time-series observations. In Stage 3 we performed a double-iteration 5$\sigma$ clipping along the time-axis for background pixels, a single iteration of 5$\sigma$ clipping along the spatial axis for each wavelength, and integration-level background subtraction using pixel columns 11--61 (excluding pixels within 10 pixels of the spectral trace) for each wavelength. Similar to the NIRCam spectra, we then performed optimal spectral extraction\citeApp{horne1986optspec} using the pixels within 4 pixels of the spectral trace using a cleaned median integration to compute our spatial profile (clipping 5$\sigma$ outliers along the time axis, smoothing with a 7-pixel wide boxcar filter, and clipping pixels that differed by more than 10$\sigma$ with respect to the spatial profile). In Stage 4, we first removed three wavelengths which exhibited excessive noise in their lightcurves. We then binned the spectra a similar binning scheme as that used by ref.\cite{dyrek2023} (47 spectral bins spanning approximately 5--12\,$\mu$m with a constant width of 0.15025\,$\mu$m); notably, however, we did not use wavelengths below 5\,$\mu$m which we expect to be significantly contaminated by light from around 3\,$\mu$m (ref.\citeApp{Kendrew2015}). We do extend our reduction out 12\,$\mu$m as was done by ref.\cite{dyrek2023} since we do not find evidence for the `shadowed region effect' reported by refs.\citeApp{bell2023_ers,bell2023arxiv} in these observations (consistent with the claim that not all observations appear to be affected by this as-yet unexplained phenomenon). Finally, as with NIRCam, we clip 4$\sigma$ outliers in the spectrally binned lightcurves compared to a cleaned version computed using a boxcar filter with a width of 20 integrations.

\subsection*{Derivation of Orbital Parameters}\label{sec:orbitParams}
All of the light curve fits to the NIRCam, HST, and MIRI data described above fix WASP-107b's orbital parameters to those derived by Murphy et al.\ (in prep.). We summarize this derivation here. 

To precisely determine WASP-107b's orbit, we simultaneously fit observations of three transits of WASP-107b from TESS, one from the Goodman High Throughput Spectrograph\citeApp{SOARGOODMANINSTRUMENT} (imaging mode, SDSS i-band) on the Southern Astrophysical Research Telescope (SOAR), one from JWST/NIRCam F210M (observed simultaneously with our NIRCam F322W2 data), one from Spitzer/IRAC with IRAC's channel 2 (publicly available from Program 13052, PI: M. Werner), and one from JWST/MIRI LRS (broadband, observation described above), as well as radial velocity (RV) observations from CORALIE\cite{Anderson17} and Keck/HIRES\cite{Piaulet21}$^\text{,}$\citeApp{KECK_HIRES}. Together, these transit observations probe a wide range of wavelengths ($\sim$0.6-12~$\mu$m) and cover a baseline spanning roughly six years, while the radial velocity observations sample WASP-107b's entire orbit. We used the default TESS pipeline reduction, reduced the SOAR/Goodman data using AstroImageJ\citeApp{ASTROIMAGEJCODE}, reduced the NIRCam/F210M data using \texttt{tshirt} (as described above), reduced the Spitzer/IRAC data using a custom photometry pipeline (described in ref.\citeApp{beatty2018_spitzerreduction}), and reduced the MIRI/LRS data using Eureka! (as described above). For the RV data, we use the measurements as tabulated in ref.\cite{Piaulet21}. 

We model each of the transits using a \texttt{batman} transit model\citeApp{batman}. For all transit observations except for SOAR/Goodman, we also fit for a background, visit-long linear trend in flux versus time. The SOAR/Goodman data exhibited significant non-linear background trends due to telluric variations during the observation, which we corrected for using a Gaussian Process model. We fit all the data described above simultaneously using MCMC sampling with \texttt{emcee}\citeApp{Foremanmackey2013}, sampling the time of conjunction, orbital period, inclination, semi-major axis, eccentricity, argument of periastron, planet-star radius ratios in each bandpass, quadratic limb-darkening coefficients in each bandpass, RV semi-amplitude and system velocity, and the slope and intercept of each visit's linear background trend. We ran this sampling for 10,500 steps, which was sufficient for each parameter to converge ($>$25x the average auto-correlation times, plus we visually inspected each MCMC walker time series). The best-fit parameter values are given in Murphy et al.\ (in prep.) and were $P$ = 5.72148722 $\pm$ $3{\times}10^{-7}$ days, $t_0$ = 2,459,958.747244 $\pm$ $8.2{\times}10^{-6}$ BJD$_{\text{TDB}}$, $i$ = 89.57 $\pm$ 0.03 degrees, $a / R_\star$ = 18.05 $\pm$ 0.1, $e$ = 0.05 $\pm$ 0.01, and $\omega$ = -2.3 $\pm$ 6.1 degrees.

\subsection*{Lightcurve Fitting}\label{sec:fitting}

\paragraph{Eureka!}\label{sec:eurekaFitting}

In \texttt{Eureka!}'s Stage 5, we fitted astrophysical and systematic noise models to our spectroscopic lightcurves. For both NIRCam filters, our systematic noise model consisted of a quadratic trend in time and a linear decorrelation as a function of the spatial position and PSF-width measured during Stage 3. For MIRI/LRS, we used a linear trend in time, an exponential ramp in time, and a linear decorrelation as a function of the spatial position and PSF-width. For all observations, our astrophysical model consisted of a \texttt{starry}\citeApp{starry} transit model with quadratic limb-darkening parameters fixed to an ATLAS model limb-darkening spectra\citeApp{kurucz1993atlas9} as computed by the Exoplanet Characterization Toolkit (ExoCTK)\citeApp{exoctk2021}. We fixed all of our orbital parameters to those of Murphy et al.\ (in prep.; see the Derivation of Orbital Parameters section above) and adopted a minimally informative prior on the planet-to-star radius ratio.

Following the recommendations of ref.\citeApp{bell2023_ers}, we removed the first 800 integrations of the MIRI/LRS observations to remove the worst part of the initial ramp in the lightcurve. We also removed integrations 475--490 from the NIRCam/F444W observations which exhibited a brief spike in noise. We removed no further integrations from the NIRCam/F322W2 observations. Finally, we also fitted a white noise multiplier to account for excess white noise. For all our observations, we used \texttt{PyMC3}'s No U-Turns Sampler\citeApp{pymc3} using two independent chains, each taking 4,000 tuning draws and then 3,000 posterior samples with a target acceptance rate of 0.85. We then validated that the chains had converged by ensuring the Gelman-Rubin statistic\citeApp{GelmanRubin1992} was at or below 1.1. The 16th, 50th, and 84th percentiles of the posterior samples were then used to estimate the best-fit values and uncertainties for each fitted parameter.

For the NIRCam/F322W2 observations, we find that the white noise is on average about 15\% above our estimate of the stellar photon-limited noise floor in each channel, while F444W exhibited only white noise only about 7\% above the estimated photon limit on average. This excess noise is likely the result of insufficiently corrected 1/$f$ noise\citeApp{schlawin2020jwstNoiseFloorI} which gets partially converted to white noise in the time-series when spectroscopically binning the lightcurves. For MIRI/LRS, we used the estimated gain of 3.1 electrons/DN from ref.\citeApp{bell2023_ers,bell2023arxiv}, as the current value of 5.5 electrons/DN used in the CRDS reference files is known to be incorrect. With this estimated gain, we find that our fitted white noise level is only about 10\% above the estimated stellar photon noise limit near 5\,$\mu$m but climbs steadily to approximately 50\% above the limit near 10\,$\mu$m and then climbs steeply to around 140\% above the limit near 12\,$\mu$m. Part of this excess long-wavelength white noise is likely due to the impact of background noise as these limits only considered the stellar photon noise. While neither of our NIRCam observations show evidence for residual red-noise in their Allan variance plots\citeApp{Allan1966}, several of our MIRI/LRS channels do show moderate amounts of residual red-noise. Following the $\beta$ error inflation method developed by ref.\citeApp{Winn2008}, we computed $\beta$, the ratio of the median value of our Allan variance plots at 15--23 minute timescales (around WASP-107b's approximately 19-minute transit ingress/egress duration) to the value expected for white noise, and multiplied our transmission spectra uncertainties by this amount. This resulted in an increase in the uncertainties by about 50\% for wavelengths $\lesssim$6.5\,$\mu$m and 0--30\% for longer wavelengths. After this, we arrived at our final 2.45--12\,$\mu$m transmission spectra.

\paragraph{Pegasus}\label{sec:pegasusFitting}

We fit the F322W2 and F444W spectroscopic lightcurves from our \texttt{Pegasus} reduction using a  \texttt{BATMAN}\citeApp{batman} transit model. We fixed most of the transit model parameters to the values measured from broadband fits to the JWST data (Murphy et al., in prep.; see the Derivation of Orbital Parameters section above), leaving only the planet-to-star radius ratio as the only free astrophysical parameter. We also simultaneously fit for a background quadratic trend across each visit, which had two associated slope parameters and a normalization factor. We did not impose a prior on any of these four parameters, and we fit each spectroscopic channel individually. For each channel, we used a quadratic limb-darkening law and fixed the limb-darkening coefficients to the values calculated from the ATLAS model\citeApp{kurucz1993atlas9} by ExoCTK\citeApp{exoctk2021} in the F322W2 and F444W bandpasses. To generate the limb-darkening coefficients we used the spectroscopic stellar properties measured in previous work\cite{Piaulet21}.

To perform the lightcurve fit in each spectroscopic channel, we performed an initial Nelder-Mead likelihood maximization followed by MCMC likelihood sampling. We initialized the MCMC chains about the maximum likelihood point estimated from the Nelder-Mead maximization. We used the \texttt{emcee}\citeApp{Foremanmackey2013} Python package to perform the MCMC sampling, using twenty walkers with a 2,000 step burn-in and a 4,000 step production run. At the end of this process we checked that the Gelman-Rubin statistics was below 1.1 for each parameter in each spectral channel to ensure the MCMC sampling had converged. 

We performed two tests to evaluate the goodness-of-fit of our transit modeling in each spectral channel. First, we confirmed that the per-point flux uncertainties the lightcurves were consistent with the standard deviation of the bestfit model residuals. Second, we calculated the Anderson-Darling statistic for the residuals to verify that they followed a Gaussian distribution. Our spectroscopic lightcurve fits did not show any evidence of non-Gaussianity in the residuals.

The estimated depth uncertainties were 1.15$\times$ the photon noise expectation in the F322W2 data and 1.06$\times$ the photon noise expectation in the F444W data.

\paragraph{tshirt}\label{sec:tshirtFitting}
We followed a similar procedure as previous NIRCam light curve fitting\cite{bell2023_methane}$^,$\citeApp{ahrer2022nircam}.
We use the \texttt{starry}\citeApp{starry} code to model the lightcurves and \texttt{pymc3}\citeApp{pymc3} to sample the posterior distributions of the fits.
We adopted uninformative priors on the stellar limb darkening with a 2 parameter quadratic law\citeApp{kipping2013limbdarkening2param}.
This allows for comparison of results with the ATLAS stellar limb darkening models.
The astrophysical \texttt{starry} model of the lightcurve is multiplied by an second order (quadratic) polynomial baseline to allow for trends from instrument and astrophysical stellar variability.
We bin the spectra in the time axis to 300 bins for faster lightcurve evaluation with \texttt{starry}.
Finally, we include a parameter for the standard deviation of the data to allow for excess noise beyond the theoretical photon and read noise. We sampled the posterior with No U-Turns sampling\citeApp{pymc3}. We used 3,000 tuning steps, 3,000 sampling steps, and 2 chains with a target acceptance rate of 90\%.

\paragraph{Comparing NIRCam Reductions}

\begin{figure*}
    \centering
    \includegraphics[width=0.75\linewidth]{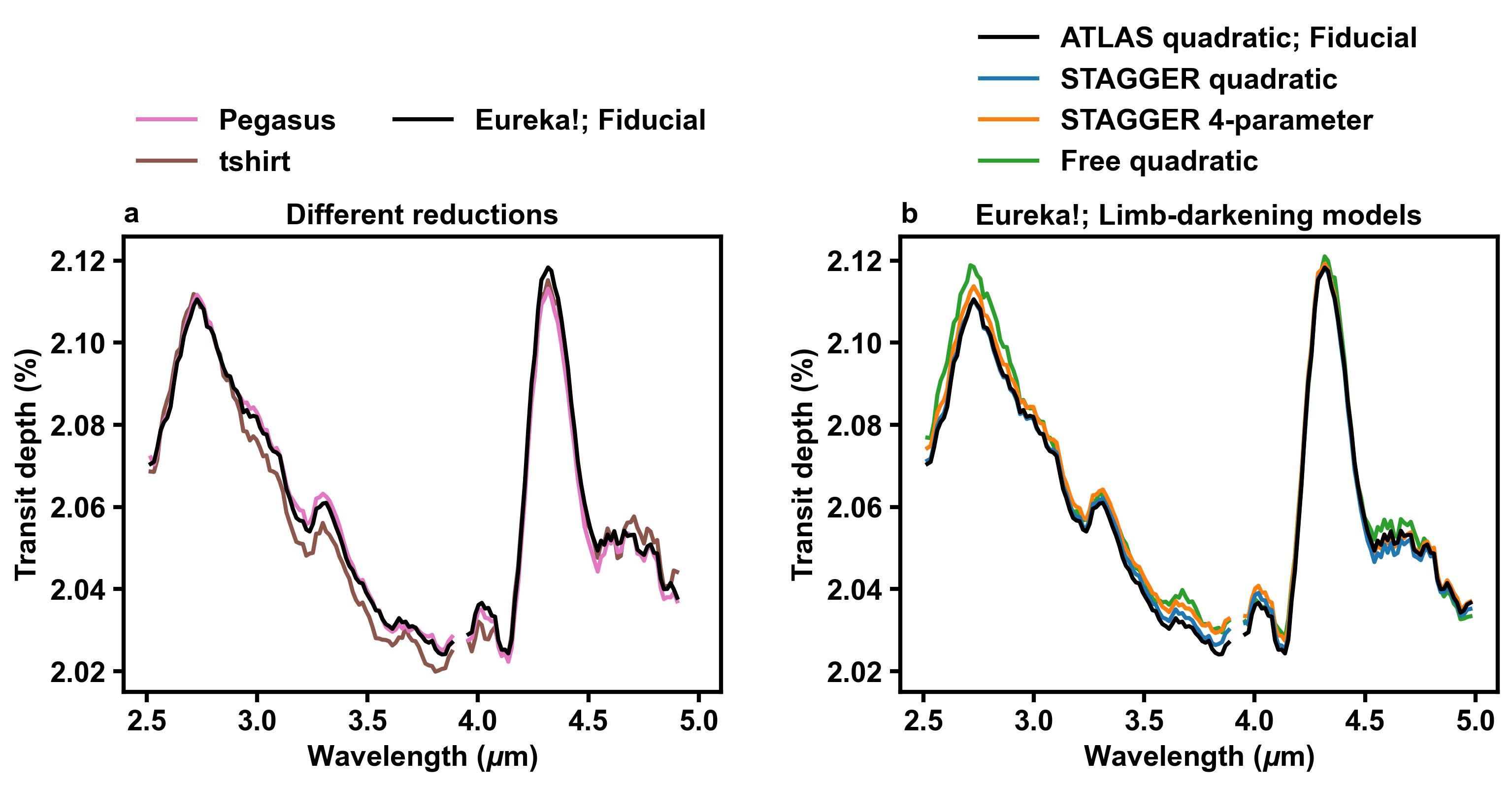}
    \caption{\textbf{Comparison of different NIRCam reductions and limb-darkening choices.} \textbf{a,} The NIRCam F322W2 and F444W transmission spectra as reduced by the \texttt{Eureka!}, \texttt{Pegasus}, and \texttt{tshirt} pipelines are shown in different colours after having been smoothed by a 4-point wide boxcar filter. Both the \texttt{Eureka!} and \texttt{Pegasus} reductions fixed limb-darkening coefficients to ATLAS model predictions, while \texttt{tshirt} freely-fit quadratic coefficients. At wavelengths longer than $\sim$2.9\,$\mu$m, \texttt{tshirt}'s F322W2 spectrum consistently falls below \texttt{Eureka!} and \texttt{Pegasus}'s spectra; this difference is caused by a combination of \texttt{tshirt}'s ROEBA method (which results in a roughly constant offset across F322W2) and \texttt{tshirt}'s free limb-darkening (which results in a small slope across F322W2). \textbf{b,} A similar plot, demonstrating the minimal impacts of varying limb-darkening choices on our final spectra produced by \texttt{Eureka!}.}
    \label{fig:nircamSpectraMethods}
\end{figure*}

A motivation for conducting three different reductions of the NIRCam data was to check the robustness of our final transmission spectrum against different choices and assumptions made during the image calibration, spectral extraction, and lightcurve fitting stages of our analyses. Generally, we find that all three reductions give the same general transmission spectrum with approximately the same depth uncertainties (Fig.~\ref{fig:nircamSpectraMain} and panel a of Extended Data Fig.~\ref{fig:nircamSpectraMethods}). The mean depth uncertainty across the entire NIRCam wavelength range is approximately 90 ppm at our constant wavelength $\Delta \lambda = 0.015$\,$\mu$m binning, which is larger than the mean point-to-point difference between the \texttt{Eureka!} and \texttt{Pegasus} spectra (58 ppm) and the \texttt{Eureka!} and \texttt{tshirt} spectra (72 ppm). A $\chi^2$ comparison of the three spectra gives $\chi^2/\textrm{dof}=0.74$ for \texttt{Eureka!} vs. \texttt{Pegasus}, and $\chi^2/\textrm{dof}=0.64$ for \texttt{Eureka!} vs. \texttt{tshirt}. The larger difference between \texttt{Eureka!} and \texttt{tshirt} is largely because \texttt{tshirt}'s F322W2 spectrum consistently falls below that of \texttt{Eureka!} and \texttt{Pegasus} at wavelengths longer than $\sim$2.893\,$\mu$m; this difference may be caused by \texttt{tshirt}'s ROEBA method since the deviation occurs right at an amplifier boundary (see panel a of Extended Data Fig.~\ref{fig:nircamSpectraMethods}). In both pipeline comparisons, the small reduced chi-squared values imply a $<$0.1$\sigma$ likelihood that the spectra are drawn from two different underlying distributions for our 172 degrees of freedom. Since the choice of reduction methods did not significantly impact our final NIRCam spectra, we chose the \texttt{Eureka!} reduction as our fiducial NIRCam spectra for all subsequent atmospheric modelling work.

To further investigate the sensitivity of our results to different limb-darkening assumptions, we also ran four separate fits with our fiducial \texttt{Eureka!} pipeline; namely, fixing our limb-darkening parameters to (1) ATLAS-predicted quadratic law coefficients computed using ExoCTK (refs.\citeApp{kurucz1993atlas9,exoctk2021}), (2) STAGGER-predicted quadratic law coefficients computed using ExoTiC-LD (refs.\citeApp{magic2015stagger,exoticld2022}), (3) STAGGER-predicted 4-parameter law coefficients computed using ExoTiC-LD (refs.\citeApp{magic2015stagger,exoticld2022}), and (4) freely fitting quadratic law coefficients using the reparameterization of ref.\citeApp{kipping2013limbdarkening2param} for efficient and physical sampling. As shown in panel b of Extended Data Figure \ref{fig:nircamSpectraMethods}, none of these spectra significantly differ from each other. In particular, we find that comparing the transmission spectrum of our fiducial analysis which used the ATLAS quadratic limb-darkening coefficients to fits with limb-darkening choices only results in a reduced chi-squared value of 0.02 for STAGGER's quadratic model, 0.09 for STAGGER 4-parameter model, and 0.28 for the fit with freely-fit reparameterized quadratic limb-darkening coefficients. In all three cases, this means that the differences between the spectra are much less than the uncertainties on the measured depths. We also further disfavour the freely-fit limb-darkening coefficients as these fitted coefficients exhibited substantially more point-to-point scatter than would be expected for a physical limb-darkening spectrum.

\paragraph{HST/WFC3}

We fit each of the twenty-nine G102 and G141 spectroscopic transit lightcurves using a \texttt{BATMAN}\citeApp{batman} transit model and standard detrending techniques for WFC3 timeseries data\citeApp{beatty2017kepler13}. As with the JWST spectroscopic data, for these fits, we fixed the transit center time, orbital period, orbital inclination, the scaled semi-major axis, the orbital eccentricity, and the argument of periastron to previously measured values (Murphy et al., in prep.; see the Derivation of Orbital Parameters section above). We also fixed the quadratic limb-darkening coefficients in each spectroscopic channel to the values estimated from the ATLAS limb-darkening model\citeApp{kurucz1993atlas9} by the Exoplanet Characterization Toolkit\citeApp{exoctk2021}. This left the planet-to-star radius ratio as the only free astrophysical parameter. We fit each spectroscopic channel individually.

All the G102 and the G141 lightcurves showed the usual `fishing-hook' systematic trend within each orbit and the background linear slope typical for WFC3 timeseries observations. We modeled this in each spectral channel as a part of the transit fitting process using an exponential ramp within each individual orbit, and a linear background trend across each visit, of the form

\begin{equation}\label{eq:3310}
F_{detrend} = (m\,t_V+n)\,\left(1-A\,e^\frac{t_O}{\tau}\right).
\end{equation}

Here, $t_O$ is the time since the start of each orbit's observations, $t_V$ is the time since the start of the visit, and $A$ and $\tau$, and $m$ and $n$ are fitting coefficients for the exponential ramp and linear trend respectively. In addition to fitting the systematics in this way, we also did not use the data from the first orbit within each visit nor the first exposure from each orbit when fitting the data.

To fit the transit lightcurve in each spectroscopic channel, we first performed a Nelder-Mead likelihood maximization to find an initial bestfit. We then used the \texttt{emcee}\citeApp{Foremanmackey2013} Python package to perform an MCMC exploration of the surrounding likelihood space to improve this bestfit estimate. In each channel we used twenty MCMC walkers, each with a 1,000 step burn-in followed by a 4,000 step production run. At the end of this we judged the MCMC to have converged by verifying that the Gelman-Rubin statistic for each parameter was below 1.1. We also checked to ensure that the median per-point flux uncertainty matched the standard deviation of the residuals to the bestfit lightcurve model in each channel. We additionally performed an Anderson-Darling test on each channel's residuals to check for non-Gaussianity. The lightcurve residuals in each channel appear well-behaved.

We note that we do not see evidence for a starspot crossing in the third orbit of the G141 transit data, as was suggested previously\cite{kreidberg2018}. Instead, the updated transit ephemeris and orbital properties we measure using the JWST data for WASP-107b serve to shift the start of transit egress slightly earlier at the time of the G141 observations. This accounts for the putative starspot feature in the earlier analyses' residuals. As a result, we do use the data from the third G141 orbit in our fitting. This slightly improves our measured depth uncertainties compared to those previous results.

\paragraph{MIRI}

\begin{figure*}
    \centering
    \includegraphics[width=.75\linewidth]{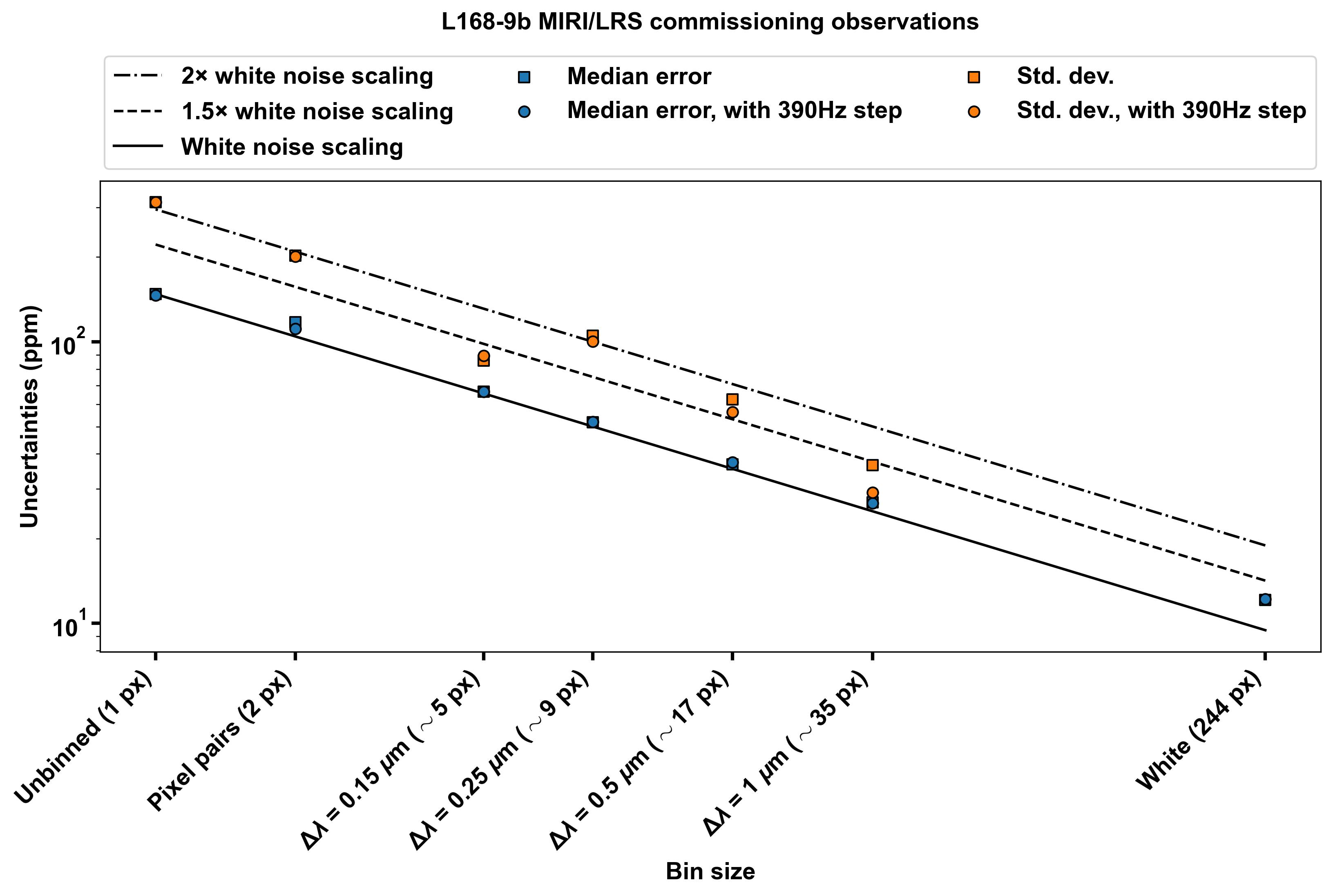}
    \caption{\textbf{A demonstration of the underestimation of error bars in the MIRI/LRS observations of L168-9b and the impact of the 390\,Hz noise removal step.} The median fitted transit depth uncertainties decrease with increasing spectral bin width as would be expected for white noise (blue symbols). Meanwhile, the standard deviation of the L168-9b transmission spectrum is around 2$\times$ the expected noise level in the spectrally unbinned data. The ratio of the standard deviation of the spectrum to the median fitted uncertainty decreases with increasing bin size with the exception of a peak around 0.25\,$\mu$m.}
    \label{fig:lrs_binning}
\end{figure*}

Unfortunately, while our 390\,Hz noise removal and GLBS steps remove structured noise from the Lomb-Scargle periodogram\citeApp{Lomb1976,Scargle1982} of the MIRI/LRS group-level data, we find that these steps ultimately have a fairly small impact on the final transmission spectra of our science target WASP-107b as well as the MIRI/LRS TSO commissioning target L168-9b. For example, Extended Data Figure \ref{fig:lrs_binning} shows that the MIRI/LRS observations of L168-9b continue to show excess noise after the 390\,Hz noise removal step has been applied. It is possible that the 390\,Hz noise removal step does eliminate excess noise for 1\,$\mu$m bin sizes, but at that coarse of a resolution there are only 7 bins and there is substantial uncertainty in the estimated standard deviation from the small sample size. In addition, with finer bin size sampling than was used by ref.\citeApp{bell2023_ers}, we are better able to understand how the excess noise varies with varying bin size. In particular, while we find that the excess noise decreases with increasing bin size (similarly to what was reported by ref.\citeApp{bell2023_ers}), there also appears to be a spike in the excess noise around a bin width of 0.25\,$\mu$m; this corresponds to an average spectral bin width of 9 pixels (although the bin width varies with wavelength) which is also the approximate period of the 390\,Hz noise. Surprisingly, while this noise spike seems to share an approximate period with the known 390\,Hz noise, the 390\,Hz noise removal step does not appear to have removed the excess noise in the final transmission spectrum of L168-9b.

\begin{figure*}
    \centering
    \includegraphics[width=0.75\linewidth]{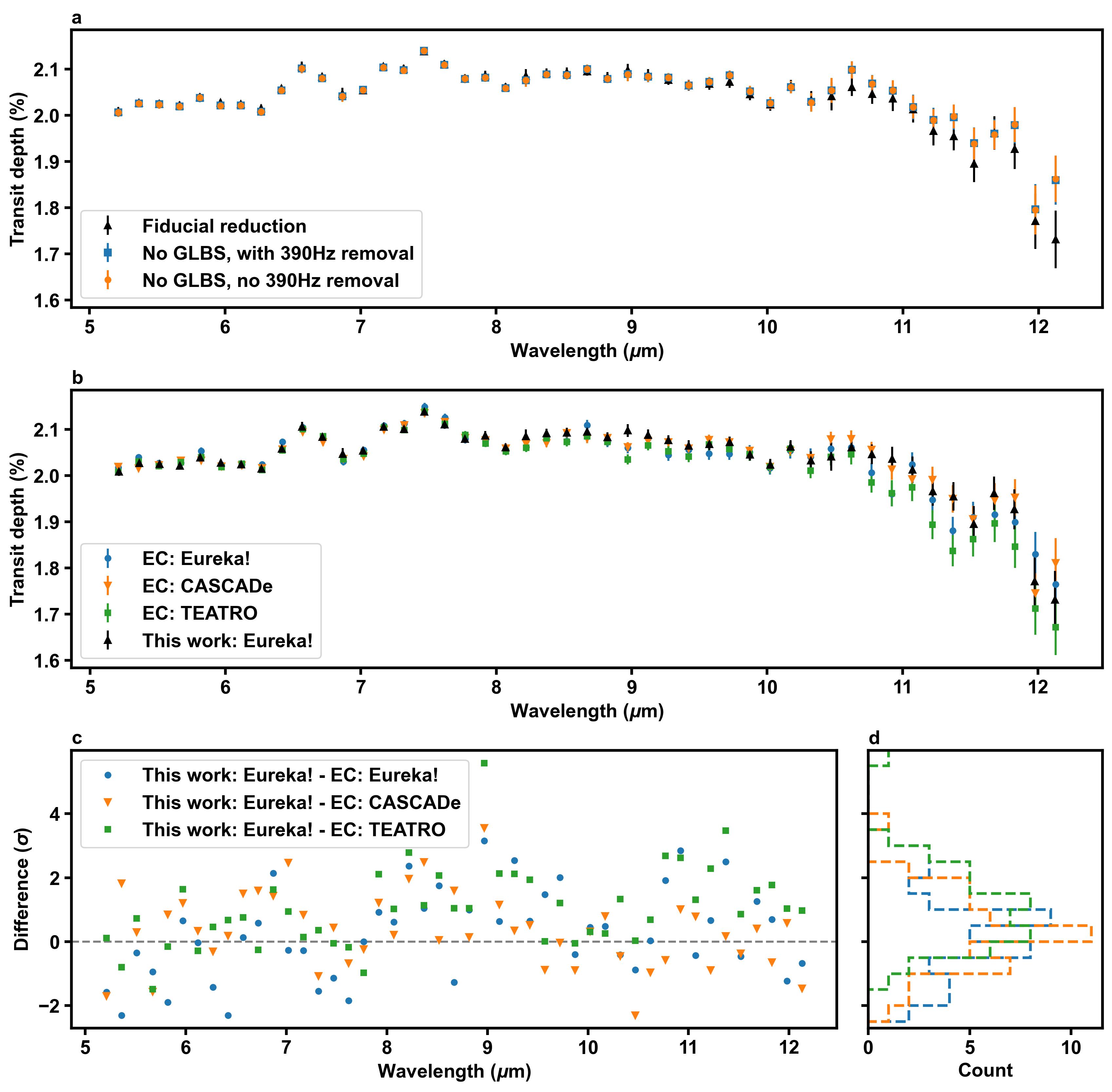}
    \caption{\textbf{A comparison of different reductions of the WASP-107b MIRI/LRS spectra.} 
    \textbf{a:} A demonstration of the impact of the 390\,Hz noise removal and GLBS steps on our reduction of the WASP-107b MIRI/LRS data. For wavelengths $>$10\,$\mu$m, the GLBS step ends up slightly reducing the final transmission, while also turning off the 390\,Hz noise removal step appears to have no further impact at any wavelength. Error bars show 1$\sigma$ uncertainties. \textbf{b:} A comparison of our fiducial reduction with the three reductions presented by ref.\cite{dyrek2023}, with 1$\sigma$ uncertainties. While there are some differences between the four different reductions, the overall shape of the transmission spectrum is robust to differences in reduction choices. \textbf{c:} The per-point differences between different pairs of reductions, which are summarized with a histogram in panel \textbf{d}. With the exception of a single \texttt{TEATRO} point, our reduction agrees with the EC's three reductions at ${\lesssim}3\sigma$. However, there is some structure to the differences between our reduction and those of the EC, with our reduction giving larger transit depths on average than the EC's from around 8--10\,$\mu$m.}\label{fig:miri_spectra}
\end{figure*}

We also see very limited impact from both the 390\,Hz noise removal and GLBS steps in the WASP-107b MIRI/LRS science observations (see Extended Data Fig.~\ref{fig:miri_spectra}). Most likely this is because there is fairly minimal curvature in the waveform over a single row (which only spans about 7\% of the first harmonic which has the largest amplitude), so the per-row background subtraction step performed on the integration-level data in Stage 3 adequately subtracts this structured noise. It is possible that this 390\,Hz noise removal procedure could still be useful to future works which seek to use separate background calibration observations taken before and/or after the science observations instead of computing the background on the science observations themselves, but at present we do not recommend the 390\,Hz noise removal procedure be applied to science observations alone due to the minimal impact and high computational cost. That said, it does not appear that our 390\,Hz noise removal procedure negatively impacts our results, and we ultimately decide to use it in our fiducial reduction. The impacts of the GLBS step are noticeable but also minor, and it is as-yet unclear whether the noise introduced by estimating the group-level background outweighs the structured noise removed by this step.

Comparing our fiducial spectrum (with the 390\,Hz noise removal and GLBS steps) to the European Consortium's (hereafter referred to as the `EC') three reductions presented by ref.\cite{dyrek2023} (see Extended Data Fig.~\ref{fig:miri_spectra}) which used the \texttt{CASCADe} and \texttt{TEATRO} pipelines, as well as the \texttt{Eureka!} pipeline, it is again clear that the overall shape of the spectrum is robust to different analysis pipelines and different reduction choices. As pointed out by ref.\cite{dyrek2023}, the transit depth differences among the EC spectra displays an increasing trend with wavelength with the \texttt{CASCADe} pipeline giving larger depths at longer wavelengths, which they attribute to different to different systematics models when fitting the lightcurves. As shown in Extended Data Figure \ref{fig:miri_spectra}, we find fairly similar conclusions with our fiducial spectrum which generally agrees well with the \texttt{CASCADe} spectrum. Looking more closely, it appears our fiducial spectrum has a slightly smaller transit depth around 7.5\,$\mu$m and a slightly larger transit depths around 7\,$\mu$m and from around 8--10\,$\mu$m compared to the three reductions of ref.\cite{dyrek2023}. This results in our fiducial spectrum having a slightly smaller bump near the $\nu_3$ SO$_2$ feature and a slightly larger bump near the $\nu_1$ SO$_2$ feature.

\subsection*{Atmospheric Retrievals}\label{sec:retrievals}

Detailed interpretation of atmospheric spectra requires comparisons with atmospheric radiative transfer models by means of statistical algorithms such as Bayesian inference methods. These data-model inference techniques are commonly known as `atmospheric retrievals' and enable constraints on the atmospheric properties of an exoplanet such as chemical composition and vertical temperature structure from an observed spectrum. The inference framework computes transmission spectra, generally on the order of millions or tens of millions per inference, from a parametric atmospheric model (that is, a `forward model') that solves line-by-line radiative transfer under hydrostatic equilibrium. These atmospheric models generally assume a plane-parallel atmosphere and include parameters that determine the chemical abundances for different chemical gases, the vertical pressure-temperature structure of the planet, and the presence of clouds/hazes as additional sources of opacity in the atmosphere. Below we present the two paradigms in atmospheric modeling employed in this study, spanning a wide range in physico-chemical assumptions.

\paragraph{1D-RCPE Grid-Retrieval}\label{sec:RCPERetrieval}

\begin{figure*}
    \centering
    \includegraphics[width=0.85\linewidth]{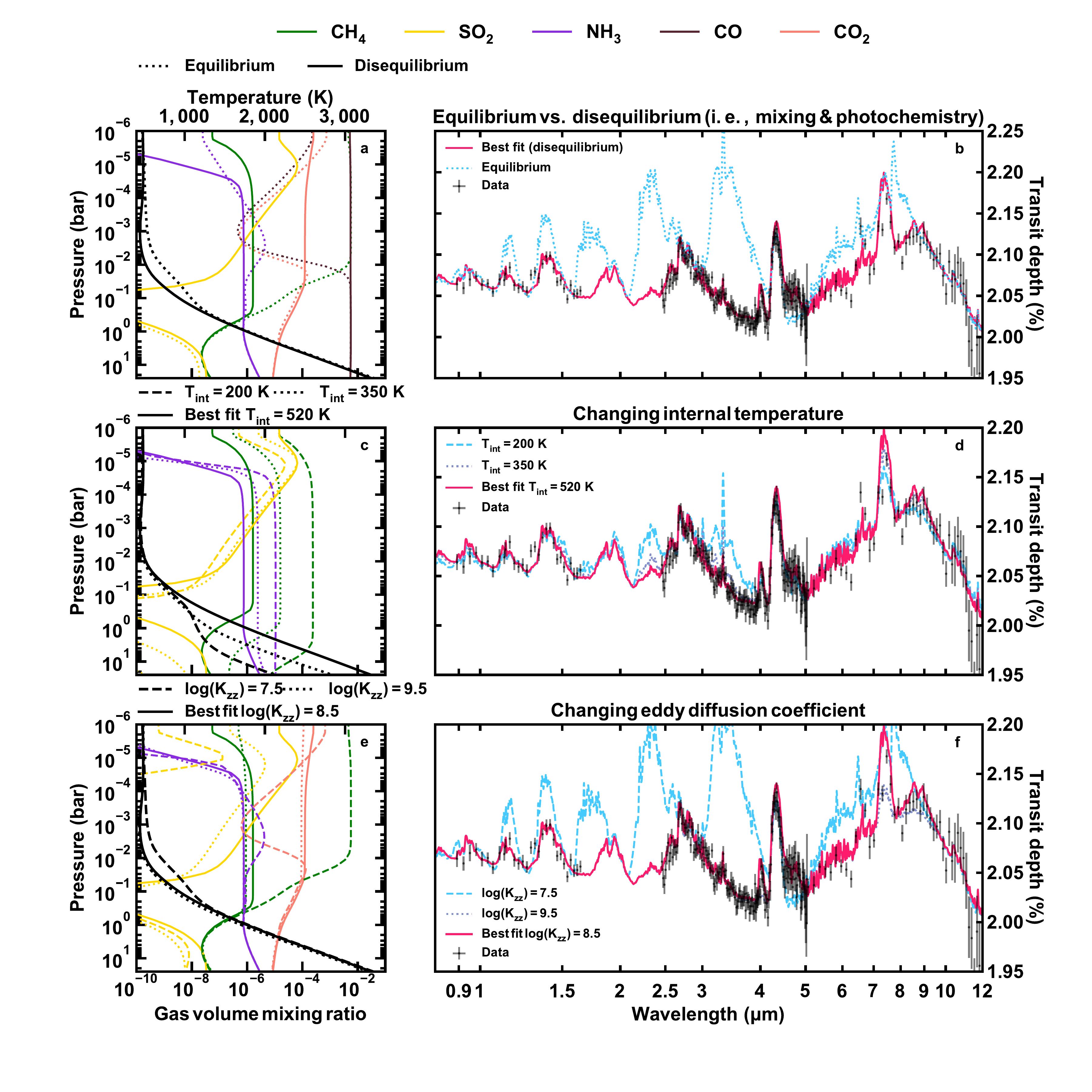}
    \caption{\textbf{Illustration of the effects of the key physical processes on the atmospheric structure (panels a, c, e) and resultant spectra (panels b, d, f)}. Panels \textbf{a} and \textbf{b} show the impact of disequilibrium chemical processes (vertical mixing and photochemistry) relative to thermochemical equilibrium. Panels \textbf{c} and \textbf{d} show the influence of changing internal temperature, while the panels \textbf{e} and \textbf{f} show the impact of changing eddy diffusion strength. We only show the vertical abundance profiles (panels \textbf{a}, \textbf{c}, \textbf{e}) for observable gases that are influenced by these effects (for instance, water is not shown as it not impacted). The observed transit spectrum is shown with grey points with 1$\sigma$ error bars in panels \textbf{b}, \textbf{d}, \textbf{f}.}
    \label{fig:perturbations}
\end{figure*}

In order to produce physically self-consistent solutions we fit the transmission spectrum with a suite of models that impose 1D-radiative convective-photochemical-equilibrium (1D-RCPE). The coupling between the radiative-convective equilibrium solver (ScCHIMERA\cite{bell2023_methane}$^,$\citeApp{Piskorz2018, Mansfield2021}) and the kinetics code (VULCAN\cite{Tsai2022}$^,$\citeApp{Tsai2017} using the H-O-C-N-S reaction list from ref.\cite{Tsai2022}) as well as additional details has been previously described in ref.\cite{bell2023_methane}. The coupled model requires as inputs the incident stellar flux spectrum (T$_{\rm eff}$=4,425\,K, log($g$)[c.g.s.]=4.63; ref.\citeApp{Husser2013}) at the top of the planetary atmosphere (including the UV for photochemistry, using the HD 85512 MUSCLES spectrum\citeApp{MusclesI,MusclesII,MusclesIII} as a proxy, which matches the effective temperature and gravity of WASP-107 well within 1\%) which can be scaled via an effective irradiation temperature, an internal temperature to set the deep adiabat, the atmospheric elemental abundances (via metallicity and a C/O as described in ref.\cite{bell2023_methane}), an eddy diffusivity, and the bulk planet properties, radius and mass. The outputs are the converged 1D temperature and gas volume mixing ratio pressure profiles that can then be `post-processed' to produce a transmission spectrum. Extended Data Figure \ref{fig:perturbations} highlights select parameter slices through the grid and the subsequent impacts on the spectra.

As this process is computationally expensive owing to the large numbers of converged models required for the Bayesian inference, we generate the grid in stages. First, we generated a course grid in metallicity (0$\le$[M/H]$\le$2 in steps of 0.5), C/O (0.1$\le$C/O$\le$0.7 in steps of 0.2), internal temperature (100$\le$T$_{\rm int}$$\le$500 in steps of 100 K), and a vertically constant eddy mixing (7$\le$log$_{10}K_{zz}$$\le$9 in steps of 1), but at a fixed effective irradiation temperature (at the planetary equilibrium temperature--738\,K).  We then perform Bayesian inference on the observed spectrum with this course grid as described in ref. \cite{bell2023_methane} (with more details specific to WASP-107b described below). Within the fitting process we also included a temperature profile offset free parameter (a simple additive shift to the whole profile) to account for possible variations in temperature between the limb and the planetary dayside which, via the scale height, can influence the feature sizes.   

The results of the course grid exercise effectively `narrowed' the plausible parameter space down at which point we generated a more finely sampled grid (below which the grid spacing had no effect). This `fine' grid was generated over 0.5$\le$[M/H]$\le$1.875 in steps of 0.125), C/O (0.05$\le$C/O$\le$0.6 in steps of 0.05), internal temperature (200$\le$T$_{\rm int}$$\le$550 in steps of 50 K), and eddy mixing (7$\le$log$_{10}K_{zz}$$\le$10 in steps of 0.5), resulting in 6,048 converged 1D-RCPE models. Based upon the temperature offset parameter from the course grid fit, this fine grid was generated at a cooler irradiation temperature (560 K) in order to more appropriately adjust for changes in the chemistry with temperature.  This finer grid (at the new fixed irradiation temperature) was then refit to the data using the same inference procedure and is what we use to inform our primary conclusions.  We also tested the effect of temperature by fitting, again, for a temperature profile offset parameter and found that it was consistent with zero at nearly 1$\sigma$ (-18$\pm$14\,K), suggesting that our choice of irradiation temperature for the fine grid resulted in a temperature profile that was consistent with the observed spectrum.  

Within the Bayesian inference procedure, as described in ref.\cite{bell2023_methane}, we also fit for, on the-fly, the planetary reference pressure and reference radius (ref.\citeApp{Welbanks2019a} which shift and stretch the planetary spectrum and also affect the zero-point for the planetary gravity with height), an offset for the MIRI spectrum relative to the NIRCam F444 spectrum, the temperature profile offset parameter (described above), an ammonia abundance enhancement factor,  a power law haze + gray cloud, and a parametric cloud profile to describe unknown cloud resonance features over the MIRI wavelengths (described more below). Including these and the 1D-RCPE grid parameters results in a total of 17 free parameters. The transmission spectra are generated with R=100,000 absorption cross sections (same line lists cited in ref.\cite{bell2023_methane}, but also now including SO$_2$, ref.\citeApp{SO2_linelist}) and fit to the data within the \texttt{PyMultiNest} routine, using a total of 500 live points and the default parameters. Gas and cloud detection significances are computed as described in refs.\cite{Welbanks2019b}$^,$\citeApp{Benneke2013}. Generally the detection of different gases in the 1D-RCPE solution corresponds to a preference for the opacity of the gas in question as a significant contribution to the spectrum, as only the opacity is removed from the inference and not from the chemistry, as described in ref.\cite{bell2023_methane}. The retrieved parameters and priors are shown in Extended Data Table \ref{tab:1DRCPE-retrieved}. 

To illustrate the sensitivity of our results to the key physical processes that enabled us to arrive at the hot interior solution, we explored perturbations to the atmospheric structure and resultant spectra, shown in Extended Data Figure \ref{fig:perturbations}. We perturb the nominal 1D-RCPE solution (given in Extended Data Table \ref{tab:1DRCPE-retrieved}) by 1) turning off mixing and photochemistry (equilibrium vs. disequilibrium, top row of Extended Data Fig.~\ref{fig:perturbations}); 2) changing the internal/effective temperature (middle row of Extended Data Fig.~\ref{fig:perturbations}); and 3) changing the strength of the eddy diffusion (bottom row of Extended Data Fig.~\ref{fig:perturbations}).  Mixing from a hot deep layer (arising from a high T$_{\rm int}$) results in a $\sim$3 order-of-magnitude depletion in the methane abundance compared to equilibrium. This has a substantial influence on the spectrum and readily explains the `muted' methane features. Equilibrium abundances are clearly ruled out by the free retrieval constraints (-6.0 $<$ log$_{10}$CH$_4$ $<$ -5.6, details below). It is also apparent that T$_{\rm int}$ values lower than $\sim$350 K and a log$_{10}K_{zz}$ lower than $\sim$8 struggle to result in the necessary reduction in the methane abundance even with mixing.

\begin{table}
    \centering
    \small
    \begin{tabular}{cl|cc}
    \hline
     & Parameter  & Value & Prior \\
    \hline
    \multirow{8}{*}{\rotatebox[origin=c]{90}{Atmospheric State}}
    &$[\textnormal{M}/\textnormal{H}]$&$1.09^{+0.17}_{-0.07}$ & (0.5, 2.0) \\
    &$\textnormal{C}/\textnormal{O}$&$0.33^{+0.06}_{-0.05}$ & (0.05, 0.60)\\
    &$T_{\textnormal{int.}}$ &$493^{+41}_{-56}$& (200, 550)\\
    &$\log_{10}\left(K_{\textnormal{zz}}\right)$&$8.62^{+0.42}_{-0.22}$& (7, 10)  \\
    &$\log_{10}(\alpha\textnormal{NH}_3)$ &$1.46^{+0.17}_{-0.19}$ & (-4, 4) \\
    &$\times R_{\textnormal{p}}$&$1.03^{+0.03}_{-0.03}$ & (0.5, 1.5) \\
    &$\log_{10}(\text{P}_\text{ref.})$  (bar)&$-3.76^{+1.5}_{-1.4}$& (-6.0, 1.5)\\
    \hline
    \multirow{9}{*}{\rotatebox[origin=c]{90}{Cloud/Haze}} 
    &$\log_{10}(\text{K}_\text{gray clouds})$ &$-38.0^{+4.8}_{-4.6}$ & (-45, -20)  \\
    &$\log_{10}(a)$ &$2.3^{+0.2}_{-0.2}$ & (-14, 14)\\
    &$\gamma$ &$-1.3^{+0.1}_{-0.1}$ & (-20, 2) \\
    &$\log_{10}(\text{K}_\text{cond.})$ &$-28.0^{+0.2}_{-0.2}$ & (-45, -20)\\
    &$\lambda_0$ &$8.2^{+0.7}_{-0.6}$ & (5, 13) \\
    &$\log_{10}(\omega)$ &$0.07^{+0.08}_{-0.05}$ & (-1, 1) \\
    &$\xi$ &$0.4^{+1.0}_{-0.9}$&  (-10, 10)\\
    &$\phi_\text{clouds and hazes}$ &$0.9^{+0.1}_{-0.1}$&  (0, 1)\\
    \hline
    &$\Delta_\text{MIRI}$ (ppm) &$282^{+49}_{-41}$& (-50, 500) \\
    \hline
    \\
    \end{tabular}
    \caption{\textbf{1D-RCPE retrieved atmospheric properties.} Retrieved parameters (median and 1$\sigma$ uncertainties) and their prior ranges are included. $\log_{10}(\alpha\textnormal{NH}_3)$ is the enhancement to the NH$_3$ abundance profile and $\Delta_\text{MIRI}$ is the offset between the MIRI observations and NIRCAM F444W. }
    \label{tab:1DRCPE-retrieved}
\end{table}

\paragraph{Cloud Parameterization}\label{sec:clouds}

There are two leading modeling paradigms for modeling clouds and hazes in exoplanetary atmospheres: microphysical models and parametric models. The former requires understanding of the nucleation, condensation, evaporation, coagulation, and transport of clouds\citeApp{Gao+21_review}. 
The later aims to capture the spectroscopic signature of these condensates and aerosols without any assumptions of the physical and chemical processes that lead to their formation and destruction, attempting to account for their presence as to unbias any inferences of other properties of interest (for example, volume mixing ratios of different gases or atmospheric metallicity, refs.\cite{Lecavelier2008a}$^\text{,}$\citeApp{Line2016a,Welbanks2022,Barstow2020a}). Some parametric models have attempted to capture the functional dependence of cloud extinction on wavelength using analytic models\citeApp{Tsiaras2018,Fisher2018} or Mie-theory\citeApp{Benneke2019a, Pinhas2018, Welbanks2021, Grant2023}. 

\begin{figure*}
    \centering
    \includegraphics[width=0.85\linewidth]{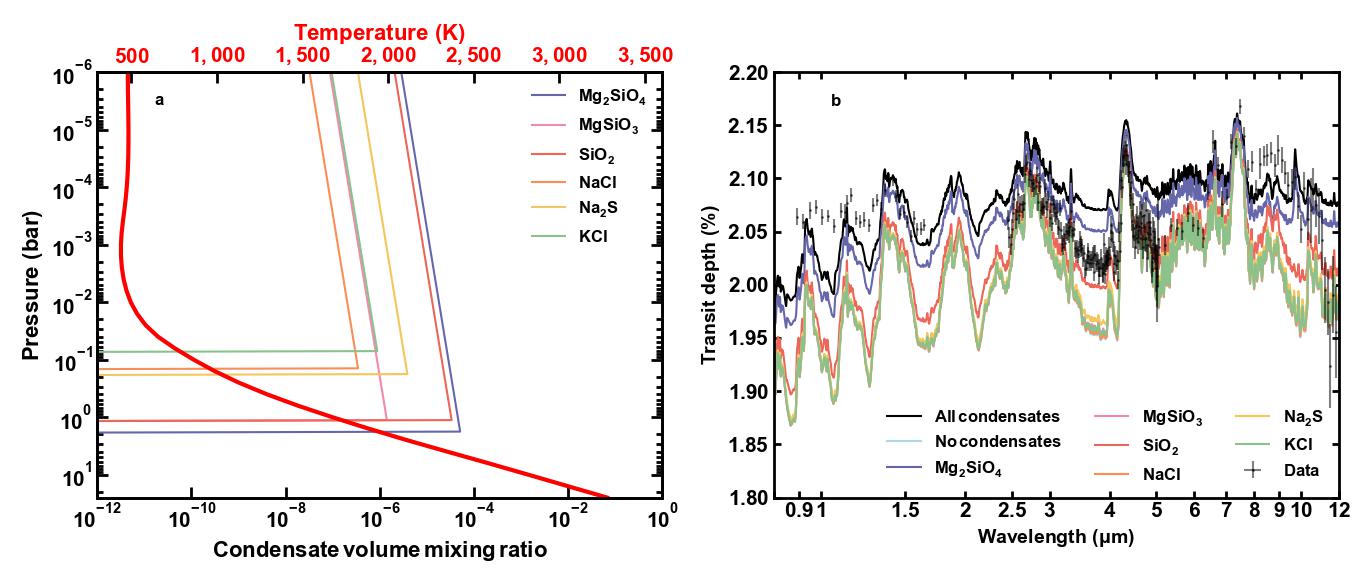}
    \caption{\textbf{Effect of self-consistent microphysical {\tt eddysed}  clouds on the spectrum.} Panel \textbf{a} shows the condensate vertical distributions for the major condensate species. The cloud bases, where the condensates first forms, all occur at or below the 100 mbar level with the silicate clouds forming at or below the 1 bar level. Panel \textbf{b} shows the resulting spectrum (with 1$\sigma$ error bars) as well as the contribution form each cloud species. }
    \label{fig:self-cons-clouds}
\end{figure*}

Of relevance to this work, one of the approaches in  ref.\cite{dyrek2023} to fit the MIRI+HST WFC3 G141 spectra of WASP-107b was to use the {\tt eddysed} microphysical cloud parameterization\citeApp{Ackerman2001, Molliere2019} for several silicate condensates. Briefly, this cloud framework assumes a balance between upwards turbulent mixing (parameterized with a vertical eddy diffusivity, $K_{zz}$) and sedimentation (parameterized with $f_{\rm sed}$) of droplets. This balance governs the particle size distribution and abundance with height in the atmosphere (larger droplets deeper towards the cloud base and smaller droplets at higher altitudes). The overall abundance of each cloud species is set by the condensate abundance at base of the cloud. The droplet abundances/profiles along with condensate/droplet optical properties (derived from Mie-Theory) are then used to compute the total cloud extinction. Specifically, they fit for a cloud base pressure, condensate abundance, $f_{\rm sed}$, and $K_{zz}$ assuming no self-consistency in the location and abundance of the condensates with the chemistry or temperature-profile.  They found that SiO$_2$ best explained the MIRI spectral shape. However, the retrieved location of the cloud base was at relatively low pressures ($\sim$mbars)--at much cooler temperatures and lower pressures than otherwise would be expected from equilibrium condensation chemistry (Extended Data Fig.~\ref{fig:self-cons-clouds}a)--and a highly compact cloud (large droplets).

We initially follow the same {\tt eddysed} cloud parameterization approach described above (assuming Mie-theory with the indices of refraction from ref.\citeApp{Wakeford2015}). While we are able to find satisfactory fits to the MIRI + HST observations (as in ref.\cite{dyrek2023}), when applied to the entire broad band spectrum with our NIRCam observations, we were unable to find satisfactory fits.

We also tested the plausibility of silicate clouds (MgSiO$_3$, SiO$_2$, and Mg$_2$SiO$_4$) and salt-sulfide clouds (Na$_2$S and KCl) in explaining the broad-band spectrum within the same {\tt eddysed}\citeApp{Ackerman2001, Mai2019} framework.  We use the 1D-RCPE atmospheric structure from an initial representative fit (T-P profile and chemistry using the nominal composition [M/H]=1.0, C/O=0.3, eddy diffusion coefficient--log$K_{zz}$=8.5, and a T$_{\rm int}$=500 K) to self-consistently determine which clouds and where they form along the temperature-profile. We varied $f_{\rm sed}$ between 0.1 (vertically extended clouds) and 1 (compact clouds). Extended Data Figure \ref{fig:self-cons-clouds} summarizes the condensate mixing ratios (panel a) and their impact on the spectrum (panel b) for an $f_{\rm sed}$=0.2.  Only the lower $f_{\rm sed}$ values ($<0.5$, more extended clouds) are able to produce a notable impact on the spectrum as higher values result in clouds that are too compact with much of the cloud opacity at too deep of pressures. However, the lower $f_{\rm sed}$ values result in smaller droplets of which produce a more narrow resonance feature than what is needed to fit the MIRI observations. It is the combination of a poor fit to the full broadband spectrum along with the in-plausibility of the silicate cloud bases to exist at the altitudes probed by the observations that instead prompts us to take a phenomenological approach to the cloud modeling. 

We introduce a new parametric treatment to capture the spectroscopic signatures of cloud particulate resonance feature. This approach is motivated by previous efforts to model optical slopes due to hazes as a scaling to Rayleigh-scattering\cite{Lecavelier2008a}. We model these concave spectroscopic signatures (see Fig.~2 of ref.\cite{dyrek2023}) of cloud condensates using a Gaussian function (that is, the composition of an exponential function and a concave quadratic function). The total extinction coefficient is given by a skewed normal distribution of form

\begin{equation}
    \kappa_\text{cond.} (\lambda)= 2\kappa_\text{opac.}\phi\left(\frac{\lambda - \lambda_0}{\omega}\right)  \Phi\left(\xi \frac{\lambda - \lambda_0}{\omega}\right)
\end{equation}

\noindent where $\kappa_\text{opac}=K_\text{cond.}n_\text{tot}$ is the extinction coefficient for an opacity free parameter $K_\text{cond.}$ and $n_\text{tot}$ is the total number density of the atmosphere; $\phi$ is the standard normal probability density function and $\Phi$ is the standard normal cumulative density functions; $\lambda_0$ is the wavelength at which the Gaussian is centered (that is, the mean of the Gaussian), $\omega$ is the scale parameter (that is, the standard deviation of the Gaussian), and $\xi$ is a shape parameter that controls the skewness of the distribution so that when $\xi=0$ the standard normal distribution is recovered. 

In both the 1D-RCPE retrievals and free retrievals we incorporate this parametric treatment alongside scalings to the Rayleigh scattering in the optical\cite{Lecavelier2008a} and optically thick gray clouds as explained in refs.\cite{bell2023_methane}, in a linear combination with a cloud-free model to account for the presence of cloud/haze inhomogeneities\citeApp{Line2016a}. Future studies may investigate the best practices for the parameterization of cloud particulate resonance features.

\paragraph{Free Retrieval}\label{sec:FreeRetrieval}

We further explore the atmospheric properties inferred from the spectrum of WASP-107b employing more flexible and agnostic forward models in a Bayesian inference procedure. These, known as `free retrievals', are methods to retrieve the chemical abundances of different gases, the vertical temperature structure, and the cloud/haze properties on the planetary atmosphere. Compared to the 1D-RCPE Grid-Retrieval explained above, these methods do not assume any physico-chemical equilibrium conditions, and instead aim to capture the atmospheric conditions directly through a series of parameters for the chemistry and physical conditions of the atmosphere without expectations of physical consistency. These more flexible approaches provide the opportunity to capture conditions that otherwise would be prohibited by self-consistent models, such as combinations of gases not considered under chemical equilibrium. Nonetheless, caution must be exercised in the interpretation of these free-retrievals as model assumptions may contribute to biased and unphysical atmospheric estimates (see ref.\citeApp{Welbanks2022} for a discussion). As with the 1D-RCPE Grid-Retrieval above, we simultaneously retrieve on the HST/WFC3 G102 and G141, JWST/NIRCam F322W2 and F444W, and JWST/MIRI observations.

We use two independent free retrieval frameworks in our analysis: Aurora\citeApp{Welbanks2021} and CHIMERA\citeApp{Line2013b,Line2016a}$^,$\cite{kreidberg2018}. We select the later as our fiducial free retrieval comparison as the radiative transfer code, sources of opacity, and model resolution is the same as in the 1D-RCPE grid-retrieval above, therefore enabling a more direct comparison. The former, Aurora, is used to consider the impact of different sources of opacity than the ones used by CHIMERA, and consider the impact of line-by-line cross-section at higher sampling resolution. We find that our detection significances and conclusions are largely independent of the framework employed, with the details of each analysis described below.

CHIMERA solves radiative transfer for a parallel-plane atmosphere, under hydrostatic equilibrium, for transmission geometry. The atmospheric model considers a one-dimensional model atmosphere spanning from ${\sim}10^{-9}$~bar to ${\sim}100$ bar, divided in 100 layers uniformly spaced in logarithmic pressure space. The vertical temperature structure is parameterized following the prescription from ref.\citeApp{Madhusudhan2009}. The model simultaneously retrieves the reference pressure and reference radius for the assumed planetary gravity of $\log_{10}(g)=2.45$~c.g.s.

The atmospheric models assumes uniform mixing ratios for H$_2$O, CH$_4$, NH$_3$, CO, CO$_2$, SO$_2$, and H$_2$S using independent free parameters for each gas' volume mixing ratio. Our choice of chemical species is motivated by those expected in exoplanetary atmospheres at these warm temperatures\cite{Moses2013, Madhusudhan2019}. As part of our initial analysis we considered the presence of PH$_3$ since this species may be expected for some substellar objects at these temperatures. Nonetheless, the observations did not place meaningful constraints on the abundance of PH$_3$ and was not considered in our fiducial run in an effort to limit the number of free parameters in our analysis. The presence of clouds and hazes is considered by utilizing the one-sector parameterization of ref.\citeApp{Welbanks2021} introducing the combined spectroscopic effect of a optically thick cloud deck with a cloud opacity $\kappa_\text{cloud}$, and hazes following an enhancement to Rayleigh-scattering\cite{Lecavelier2008a} in a linear combination with a cloud-free atmosphere\citeApp{Line2016a}. Furthermore, we incorporate the effect of unknown cloud resonances (for example, wavelength dependent condensates) as explained above.

The Bayesian inference is performed using nested sampling\citeApp{Skilling2004, Feroz2009}, via \texttt{PyMultiNest}\cite{Buchner2014}$^\text{,}$\citeApp{Feroz2009,Feroz2019}, using 500 live points in the sampling. Each forward model in the sampling was calculated using line-by-line opacity sampling at a spectral resolution of 100,000 and then binned to the resolution of the observations. The line lists considered remain the same as in the 1D-RCPE analysis described above. In total, the sampling is performed over 24 parameters: 7 molecular gases, 6 for the pressure-temperature structure of the planet, 1 for the reference pressure, 1 for the reference radius, 8 for the presence of inhomogeneous clouds and hazes, and 1 for an offset between the JWST MIRI observations and the JWST NIRCam F444W observations. 

Our second free-retrieval with Aurora\citeApp{Welbanks2021} follows largely the same model setup as the one explained above with CHIMERA. A detailed description of Aurora's transmission modelling approach is available in ref.\citeApp{Welbanks2021}. The main differences relative to the retrieval with CHIMERA are: 1) different opacity sources for H$_2$O, CO$_2$, NH$_3$, CH$_4$, and H$_2$S obtained from refs.\citeApp{Rothman2010,Yurchenko2014,Yurchenko2011,Underwood2016}; 2) the use of a free parameter P$_{\rm{cloud}}$ to account for optically thick clouds instead of $\kappa_{\rm{cloud}}$parameter; and 3) computing the forward models using line-by-line cross section sampling at a spectral resolution of 20,000 instead of 100,000. The retrieved atmospheric properties are generally consistent within the the CHIMERA retrieval at 68$\%$ confidence, and with predictions from the 1D RCPE grid retrieval. The results from this retrieval are also included in Table \ref{tab:free-retrieved}. Supplementary Information Figure 2 shows the equivalent to Figure \ref{fig:grid_v_free} with the Aurora constraints, and highlights that while the inferences are consistent, the use of lower spectral resolution results in wider constraints in this case.

\begin{figure}
    \centering
    \includegraphics[width=\linewidth]{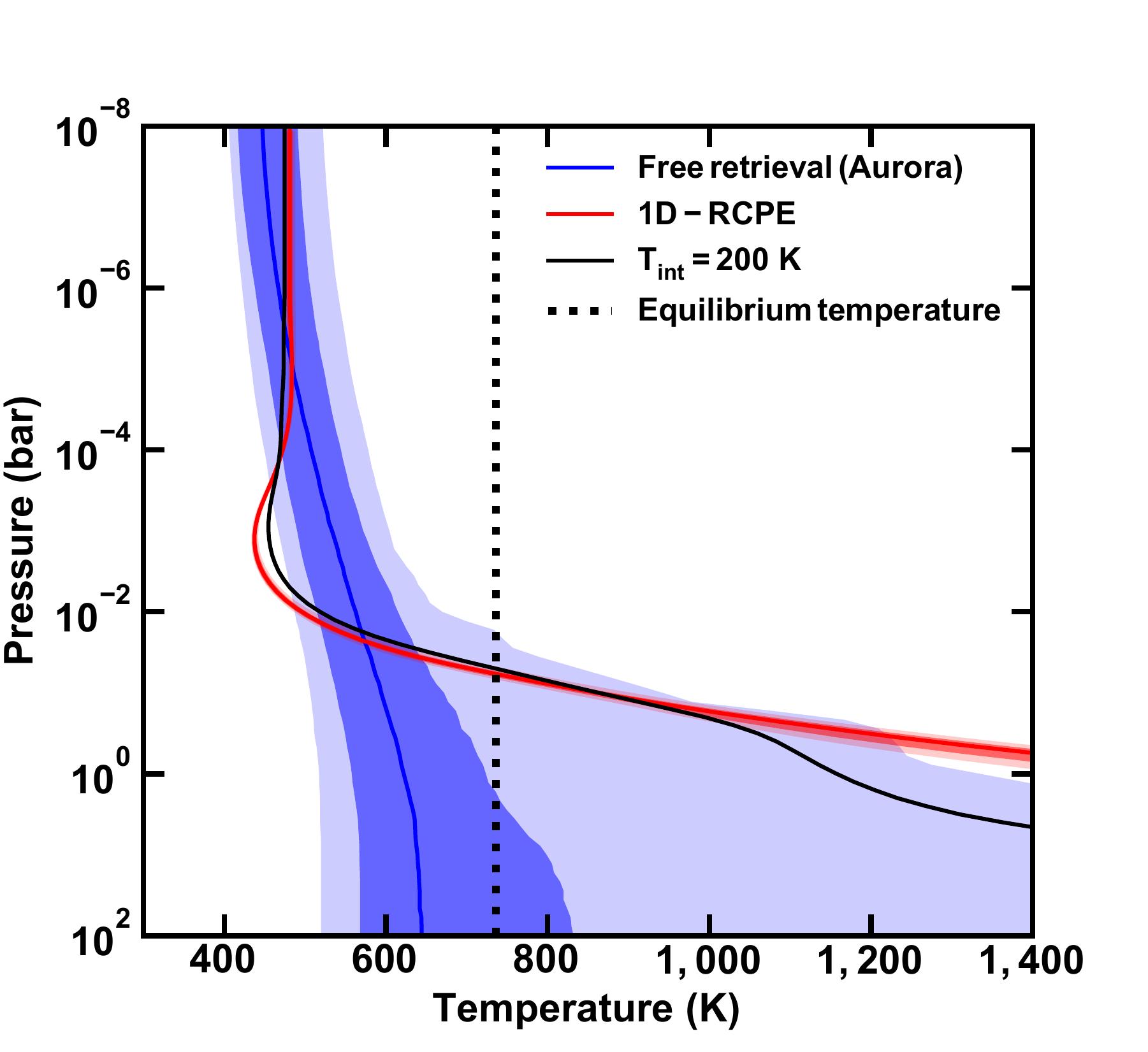}
    \caption{\textbf{Comparison of retrieved pressure-temperature structure.} The retrieved vertical temperature structures (median, 1$\sigma$, and 2$\sigma$) from the free-retrieval (Aurora, blue) and the 1D-RCPE (red) are generally in good agreement within their 68\% confidence intervals. The dotted line shows the equilibrium temperature of the planet. The observations generally probe pressures lower than $\sim$100 mbar and as low as $\sim$few $\times 10^{-5}$ bar. We include in black the T$_\text{int}=200$~K profile from Extended Data Figure \ref{fig:perturbations} as an example of lower internal temperatures. }
    \label{fig:PT_comparison}
\end{figure}

Supplementary Information Figure 1 shows the retrieved transmission spectrum and the contributions of the detected gases in WASP-107b's atmosphere as inferred with Aurora. In both cases, the retrieved volume mixing ratios from the free-retrievals are consistent with the gas profiles from the 1D-RCPE inferences, as shown by the posterior distributions from the free retrieval and gas samples from the 1D-RCPE in Figure \ref{fig:grid_v_free} and Supplementary Information Figure 2. Furthermore, the retrieved vertical pressure-temperature structure of the planet is consistent within the fully parametric free-retrieval and the self-consistent 1D-RCPE as shown in Extended Data Figure \ref{fig:PT_comparison}. The free retrieval with Aurora finds an offset between the JWST MIRI and JWST NIRCam F444W observations of $282 ^{+49}_{-41}$ ppm. A complete summary of the retrieved parameters and their priors is shown in Extended Data Table \ref{tab:free-retrieved}. 

\begin{table*}  
    \centering
    \small
    \begin{tabular}{cl|ccc}
    \hline
     & Parameter  & CHIMERA & Aurora & Prior\\
    \hline
    \multirow{6}{*}{\rotatebox[origin=c]{90}{Chemical Species}}
    &$\log_{10} \left(X_{\textnormal{H}_2\textnormal{O}}\right)$&$-2.1 ^{+ 0.2 }_{- 0.3 }$&$-2.6 ^{+ 0.7 }_{- 0.6 }$&($-12.0,-0.3$) \\
    &$\log_{10}\left(X_{\textnormal{CH}_4}\right)$&$-5.8 ^{+ 0.2 }_{- 0.2 }$&$-6.1 ^{+ 0.5 }_{- 0.4 }$ &($-12.0,-0.3$) \\
    &$\log_{10}\left(X_{\textnormal{NH}_3}\right)$ &$-5.0 ^{+ 0.2 }_{- 0.2 }$&$-5.1 ^{+ 0.5 }_{- 0.5 }$ &($-12.0,-0.3$) \\
    &$\log_{10}\left(X_{\textnormal{CO}}\right)$&$-1.9 ^{+ 0.2 }_{- 0.2 }$&$-3.0 ^{+ 0.7 }_{- 0.7 }$ &($-12.0,-0.3$) \\
    &$\log_{10}\left(X_{\textnormal{CO}_2}\right)$&$-3.9 ^{+ 0.3 }_{- 0.3 }$&$-4.4 ^{+ 0.8 }_{- 0.7 }$&($-12.0,-0.3$)  \\
    &$\log_{10}\left(X_{\textnormal{SO}_2}\right)$&$-5.2 ^{+ 0.2 }_{- 0.2 }$& $ -5.7 ^{+ 0.4 }_{- 0.4 }$&($-12.0,-0.3$)  \\
    &$\log_{10}\left(X_{\textnormal{H}_2\textnormal{S}}\right)$&$-8.5 ^{+ 2.2 }_{- 2.2 }$& $-8.6 ^{+ 2.1 }_{- 2.2 }$ &($-12.0,-0.3$) \\
    \hline
    \multirow{6}{*}{\rotatebox[origin=c]{90}{P-T}} 
    &T$_{0}$ (K) &$525 ^{+ 26 }_{- 31 } $&$447 ^{+ 42 }_{- 30 } $ & (400, 900) \\
    & $\alpha_1$  (K$^{-1/2}$)&$1.6 ^{+ 0.3 }_{- 0.3}$&$1.6 ^{+ 0.3 }_{- 0.3}$  & (0.02, 2.0) \\
    &$\alpha_2$  (K$^{-1/2}$)  &$1.2 ^{+ 0.6 }_{- 0.6 }$&$1.2 ^{+ 0.5 }_{- 0.6 }$  & (0.02, 2.0)\\
    &$\log_{10}(\text{P}_1$) (bar) &$-2.4 ^{+ 1.6 }_{- 1.9 }$&$-2.0 ^{+ 1.7 }_{- 1.9 }$ & (-9.0, 2.0) \\
    &$\log_{10}(\text{P}_2$)(bar) &$-5.8 ^{+ 2.0 }_{- 1.8 }$&$-5.8 ^{+ 2.5 }_{- 2.1 }$ & (-9.0, 2.0)\\
    &$\log_{10}(\text{P}_3$) (bar) &$-0.1 ^{+ 0.9 }_{- 1.0 }$&$0.4 ^{+ 1.1 }_{- 1.3 }$ & (-2.0, 2.0) \\
    \hline
    &$\log_{10}(\text{P}_\text{ref.}$)  (bar)&$-4.9 ^{+ 2.4 }_{- 2.0 } $&$-7.3 ^{+ 1.3 }_{- 1.0 } $ & (-9.0, 2.0)\\
    &$\text{R}_\text{p}$  ($\text{R}_\text{Jup.}$ )&$0.98 ^{+ 0.04 }_{- 0.05}$&$1.01 ^{+ 0.02 }_{- 0.03}$ & (0.66, 1.22)\\
    \hline
    \multirow{9}{*}{\rotatebox[origin=c]{90}{Cloud/Haze}} 
    &$\log_{10}(a)$&$2.5 ^{+ 0.2 }_{- 0.2 }$&$1.7 ^{+ 0.5 }_{- 0.4 }$ & (-4, 10)\\
    &$\gamma$ &$-1.0 ^{+ 0.1 }_{- 0.1 }$&$-1.2 ^{+ 0.1 }_{- 0.2 }$ & (-20, 2) \\
    &$\log_{10}(\text{K}_\text{cond.})$ &$-27.9 ^{+ 0.2 }_{- 0.2 }$&$-28.4 ^{+ 0.5 }_{- 0.3 }$ & (-50, -20)\\
    &$\lambda_0$ &$7.7 ^{+ 0.5 }_{- 0.4 }$&$7.4 ^{+ 0.5 }_{- 0.3 }$ & (5.5, 12.0) \\
    &$\log_{10}(\omega)$ &$0.1 ^{+ 0.1 }_{- 0.1 }$&$0.2 ^{+ 0.1 }_{- 0.1 }$ & (-1, 1) \\
    &$\xi$ &$1.1 ^{+ 0.9 }_{- 0.7 }$&$2.3 ^{+ 1.5 }_{- 1.2 }$&  (-10, 10)\\
    &$\log_{10}(\text{P}_\text{cloud})$ (bar) &N/A&$-0.1 ^{+ 1.3 }_{- 1.3 }$ &(-9.0, 2.0) \\
    &$\log_{10}(\kappa_\text{cloud})$ (bar) &$-42.8 ^{+ 8.1 }_{- 7.4 }$&N/A &(-9.0, 2.0) \\
    &$\phi_\text{clouds and hazes}$ &$0.9 ^{+ 0.1 }_{- 0.1 }$&$0.9 ^{+ 0.1 }_{- 0.1 }$&  (0, 1)\\
    \hline
    &$\Delta_\text{MIRI}$ (ppm) &$ 253^{+ 41}_{- 39 }$&$ 285^{+ 37}_{- 36 }$& (-50, 500) \\
    \hline
    \\
    \end{tabular}
    \caption{\textbf{Free-retrieval retrieved atmospheric properties.} Retrieved parameters (median and 1$\sigma$ uncertainties) and their prior ranges are included.}
    \label{tab:free-retrieved}
\end{table*}

Both free retrievals confirm the detections of H$_2$O, CH$_4$, NH$_3$, CO, CO$_2$, and SO$_2$ with detection significances of $5\sigma$ or greater. The reported detection significances in the main text result from the model comparisons with CHIMERA, while the detections with Aurora remain equally significant at $5\sigma$ or greater. Similarly, we find that the fiducial model with optically thick clouds, Rayleigh-enhancement hazes, and wavelength dependent condensates is preferred over a cloud-free atmosphere at $26\sigma$. Including wavelength dependent condensates is preferred at $13\sigma$ over just including optically thick clouds and Rayleigh-enhancement hazes. We note that the term `detection significance' refer to a model preference between a reference model considering all gases and nested models for which each individual species is removed in turn\citeApp{Benneke2013, Welbanks2021}. As such, any quoted detection significance is dependent on the choice of models and choice of model priors. Moreover, the conversion between differences in Bayesian evidence and `sigma detections' can result in extremely large values (for example, $>10\sigma$) that are difficult to interpret (see ref.\citeApp{Welbanks2023} for a discussion). For WASP-107b, the agreement in retrieved molecular abundances and confirmation of the strong evidence for the detection of these species regardless of model assumptions (for example, free abundances vs. radiative convective-photochemical-equilibrium) confirms the robustness of the interpretation of the planet's spectrum.

\paragraph{Robustness of Ammonia Detection}\label{sec:NH3Detection}

\begin{figure*}
    \centering
    \includegraphics[width=0.75\linewidth]{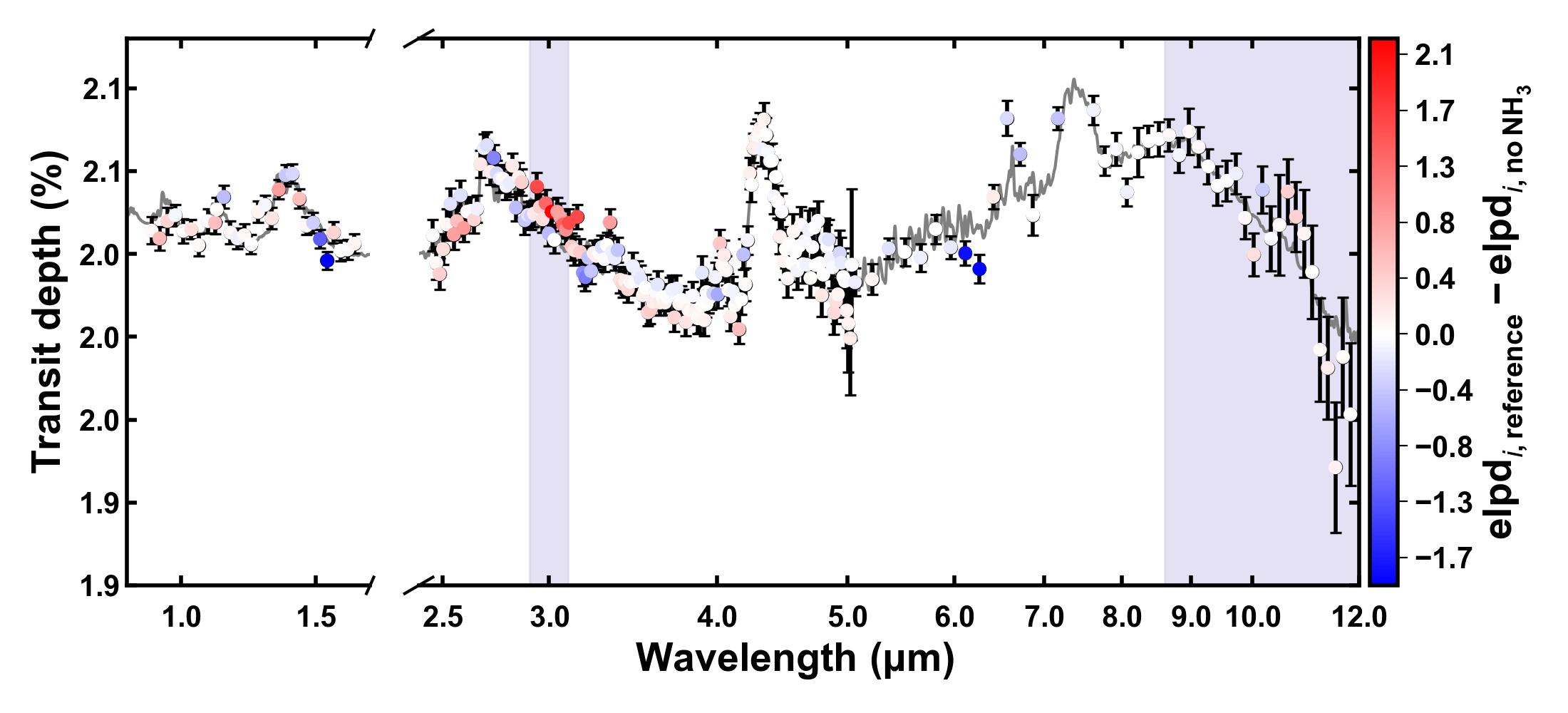}
    \caption{\textbf{The detection of NH$_3$ in WASP-107b's transmission spectrum is driven by NIRCAM F322W2 observations.} The data is color coded by the point-wise difference in expected log posterior predictive density between the reference 1D-RCPE model and the model without NH$_3$ absorption. Best fit 1D-RCPE model (R=300) is show in gray. Redder data points (larger positive $\Delta {\rm elpd}$ score) are better explained by the reference model including NH$_3$ absorption. While points with positive $\Delta {\rm elpd}$ are present throughout the entire transmission spectrum, indicating that NH$_3$ improves the model performance at all wavelengths, the highest scoring points are localized at ${\sim}3\mu$m where strong NH$_3$ absorption is visible. The purple shaded regions show areas where the cross-section of NH$_3$ contributes to 50\% or more of the cross-sections detected in WASP-107b, and with visually prominent features. Three data points in the MIRI observations (7.017\,$\mu$m, 7.318\,$\mu$m, 7.468\,$\mu$m) are not included in the analysis as their associated Pareto $k$ value exceeded 0.7. Data points are shown with 1$\sigma$ error bars.}\label{fig:loo_analysis}
\end{figure*}

To provide scrutiny beyond using Bayesian evidence model selection for the NH$_{3}$ detection, we perform Leave-One-Out Cross-Validation (LOO-CV; refs.\citeApp{Vehtari2012, Vehtari2017, Vehtari2024}) on the models with and without NH$_{3}$ from the 1D-RCPE grid retrievals. In LOO-CV, a model is trained on the dataset with one data point left out and then scored according to how it can predict the left out data point (that is, by calculating the expected log predicted density of the left out data point - elpd). This procedure is repeated for all data points allowing the out-of-sample predictive performance of a model to be estimated. The LOO-CV scores for each data point can then be compared between two models highlighting where one model out performs the other\citeApp{Welbanks2023, McGill2023, Challener2023} and consequently, in this case, which wavelengths and instruments drive the NH$_{3}$ detection. Extended Data Figure \ref{fig:loo_analysis} shows the difference in elpd ($\Delta$elpd) per data point in the spectrum of WASP-107b for the reference 1D-RCPE model and the model without NH$_3$ absorption.

The Bayesian model comparison finds a 6$\sigma$ detection (that is, model preference) for the model including NH$_3$ absorption. By performing a LOO-CV analysis of these models, we can determine which data points drive this model preference. We find the inclusion of NH$_3$ improved the model performance at all wavelengths (for example, there are points with positive $\Delta {\rm elpd}$ at all wavelengths), with a localized region of high performance data points near $3\,\mu$m. Indeed, the points with the highest preference for the model with NH$_3$, those with $\Delta {\rm elpd}>1$, are all between $2.9\,\mu$m and $3.2\,\mu$m and part of the NIRCam F322W2 observations. After NIRCAM F322W2, the next instrument with with positive $\Delta {\rm elpd}$ scores is HST-WFC3 G141 (9$^{\rm th}$ highest scoring point), followed by HST-WFC3 G102 (14$^{\rm th}$ highest scoring point) and NIRCAM F444W (seventeenth highest scoring point). MIRI's highest scoring point is at 10.17\,$\mu$m and corresponds to the 25$^{\rm th}$ highest scoring point with a $\Delta {\rm elpd}\sim0.4$. 

We proceed to total the difference in LOO-CV scores (that is, \mbox{sum($\Delta {\rm elpd})$}) between the models with and without NH$_3$ over regions where the NH$_3$ cross-section is dominant (that is, where NH$_3$ contributes $>50\%$ of the total cross-section). We find that the density of the increased predictive performance (that is, \mbox{sum($\Delta {\rm elpd})$/\# points}) is higher, and more than double the value, in regions where NH$_3$ is dominant than in regions where NH$_3$ is not the dominant cross-section. This confirms that the detection of NH$_3$ is significantly improved by the data where significant absorption by the gas is expected.

We also compare two regions in the spectrum where the spectroscopic signatures of NH$_3$ are visually prominent, these are features at ${\sim}3\,\mu$m with NIRCam F322W2 and ${\sim}10\,\mu$m with MIRI. These two features are part of regions where  NH$_3$ contributes $>50\%$ of the total cross-section (see purple shaded regions in Extended Data Fig.~\ref{fig:loo_analysis}). The region between 2.9\,$\mu$m and $3.1\,\mu$m covered by NIRCam F322W2 has an order of magnitude greater density in the increased predictive performance of NH$_3$ than the 8.6\,$\mu$m to 12\,$\mu$m region covered by MIRI. Our analysis finds that the detection of NH$_3$ is more strongly driven by the NIRCam observations rather than the MIRI observations.

The NIRCam F322W2 provide the information necessary to confirm the tentative detection (${\sim}2$--3$\sigma$) of NH$_3$ suggested by ref.\cite{dyrek2023} using MIRI data alone. Our LOO-CV analysis suggests that definitive detections of NH$_3$ require resolving the strong spectroscopic feature at ${\sim}3.0\mu$m, and highlight an important advantage of NIRCam over other instruments on JWST that do not have the wavelength coverage or throughput necessary. While NH$_3$ has a strong spectral feature at ${\sim}10\,\mu$m, and the gas' cross-section is dominant at $>8.6\,\mu$m, the presence of strong cloud resonance features in the infrared obfuscates the observational ability to strongly detect the gas. 

\paragraph{Stellar SED Fitting and Absolute Radii}\label{sec:SEDandRadius}

We used a set of catalog magnitudes for the WASP-107 system to fit a model SED to the stellar emission. In conjunction with the Gaia\citeApp{GaiaDR3} parallax for the system, this allowed us to estimate a stellar radius for the star WASP-107.

For our SED fits, we used catalog 2MASS JHK\citeApp{2mass} and AllWISE W1 and W2\citeApp{wise} magnitudes. Our SED model used four different physical parameters: the stellar effective temperature, the stellar radius, the amount of visual extinction to WASP-107, and the system's parallax. We assumed $\log(g)=4.5$ and $[\mathrm{Fe}/\mathrm{H}]=0.0$ for the star WASP-107, which are both within 0.1 dex of the surface gravity and metallicity measured previously.\cite{Piaulet21} In practice, the precise values of $\log(g)$ and $[\mathrm{Fe}/\mathrm{H}]$ do not significantly affect the results from the SED fit.\citeApp{stevens2018}

We imposed a Gaussian prior on T$_\text{eff.}$ based on previous spectroscopic measurements\cite{Piaulet21} of T$_\text{eff.}=4425\pm70$\,K with the associated $1\sigma$ uncertainties as the prior width. Similarly, we imposed a Gaussian prior on the parallax to the system using the Gaia DR3\citeApp{GaiaDR3} parallax of $\pi=15.528\pm0.026$\,mas. For the amount of visual extinction to WASP-107 we imposed a prior of $A_V=0.03\pm0.01$, which comes from measurements\citeApp{sf2011dustmap} of the excess reddening towards WASP-107 of $E(B-V)=0.028\pm0.01$, and assuming $R_V=3.1$.

To model the SED, we used BHAC15 spectra\citeApp{bhac15} for the stellar SED. We computed a grid of surface luminosity magnitudes, corresponding to the bandpasses of the catalog magnitudes, for a range of T$_\text{eff.}$ values. Since the BHAC15 models step by 200\,K in T$_\text{eff}$, we used cubic spline interpolation to estimate model magnitudes in between the points provided by the model atmospheres. We then scaled the interpolated surface magnitudes for the star by $R_*/d$ -- where $d$ is the distance to the star -- to determine the apparent bolometric flux of the SED at Earth. We then applied a simple $R=3.1$ extinction law scaled from the value of $A_V$, to determine the extincted bolometric flux of the SED model.

This stellar SED fitting allowed us to measure the radius of the star WASP-107 to be $R_*=0.67\pm0.01\,R_\odot$. We then used the mean value of $R_{\rm p}/R_*$ as measured in our joint broadband transit fits to estimate the planetary radius of WASP-107b to be $R_{\rm p}=0.939\pm0.019\,R_{\rm J}$.

\subsubsection*{Data Availability}
The NIRCam data used in this paper are associated with JWST GTO program 1185 (PI Greene; observations 8 and 9) and will be publicly available from the Mikulski Archive for Space Telescopes (MAST; \url{https://mast.stsci.edu}) at the end of their one-year exclusive access period. The MIRI data used in this paper are from JWST GTO program 1280 (PI Lagage; observation 1) and will also be publicly available on MAST at the end of their proprietary period. Additional intermediate results from this work are archived on Zenodo at \url{https://doi.org/10.5281/zenodo.10780448} (ref.\citeApp{welbanks2024_wasp107b_zenodo}).\\

\subsubsection*{Code Availability}
We used the following codes to reduce and fit the JWST data: STScI's JWST Calibration pipeline\citeApp{jwst_v1.10.2}, \texttt{Eureka!}\cite{bell2022}, \texttt{tshirt}\cite{tshirt:2022}, starry\citeApp{starry}, PyMC3\citeApp{pymc3}, numpy\citeApp{numpy}, astropy\citeApp{astropy2013,astropy2018}, scipy\citeApp{scipy}, and matplotlib\citeApp{matplotlib}.\\

\bibliographystyleApp{sn-standardnature}

\bmhead{Acknowledgments} 

L.W.\ acknowledges support for this work provided by NASA through the NASA Hubble Fellowship grant \#HST-HF2-51496.001-A awarded by the Space Telescope Science Institute, which is operated by the Association of Universities for Research in Astronomy, Inc., for NASA, under contract NAS5-26555. T.P.G.\ and T.J.B.\ acknowledge support from NASA JWST WBSs 411672.07.04.01.02 and 411672.07.05.05.03.02. P.M.\ acknowledges that this work was performed under the auspices of the U.S. Department of Energy by Lawrence Livermore National Laboratory under Contract DE-AC52-07NA27344. The document number is LLNL-JRNL-859050. M.R.L.\ acknowledges NASA XRP award 80NSSC19K0446 and STScI grant HST-AR-16139. M.R.L.\ and L.W.\ acknowledge Research Computing at Arizona State University for providing HPC and storage resources that have significantly contributed to the research results reported within this manuscript. L.W.\ thanks Michiel Min for meaningful conversations. K.O.\ acknowledges support from JSPS KAKENHI Grant Number JP23K19072. M.M.\ acknowledges funding from NASA Goddard Spaceflight Center via NASA contract NAS5-02105. A.D.\ and P-O.L.\ acknowledge funding support from CNES. We thank Marcia Rieke for allocating the NIRCam time for this program.

\bmhead{Author Contribution}

L.W.\ led the modeling analysis effort, performed the atmospheric modeling using free-retrievals and 1D-RCPE models, contributed to the theoretical analysis/interpretation of the observations, led the cross-validation analysis, and led the writing of the manuscript. T.J.B.\ contributed the fiducial \texttt{Eureka!} analyses of the NIRCam and MIRI observations, verified the observing parameters of the NIRCam observations, and contributed to the text. T.G.B.\ contributed the \texttt{Pegasus} analyses of the HST and NIRCam observations, verified the observing parameters of the NIRCam observations, performed the SED fitting and radius estimation, performed the tidal heating analysis, and contributed to the text. M.R.L.\ performed the 1D-RCPE simulations, wrote introductory text and methods, and helped sculpt the direction of the manuscript. K.O.\ provided comments on the manuscript and helped with interpreting the NH3 detection. J.J.F.\ contributed to the planet structure models, contributed to the interpretation of the internal temperature of the planet, and contributed to the text. E.S.\ simulated the planet and retrievals before launch, help planned the observation specifications, did the \texttt{tshirt} reduction and light curve fitting, and contributed to the text. T.P.G.\ selected the planet for observation, designed the observational program, directed some of the analysis, and commented on the manuscript. E.R.\ provided feedback on the modeling interpretation and provided comments on the manuscript. P.M.\ contributed the cross-validation analysis and to the text. V.P.\ provided feedback throughout the project and helped with the vertical mixing discussion. Y.T.\ contributed to the planet structure models. M.M.\ contributed to deriving the planet's orbital parameters and feedback on the text. I.E., S.M., L.S.W., and K.E.A. provided useful discussions throughout the manuscript process. P-O.L. designed the MIRI observational program and provided its data. A.D. contributed to the MIRI data analysis.\\

\noindent \textbf{Competing Interests Statement} The authors declare no competing interests.\\

\noindent\textbf{Additional information}\newline
\textbf{Correspondence and requests for materials} should be addressed to \href{Luis Welbanks}{mailto:luis.welbanks@asu.edu}.
\newline
\textbf{Reprints and permissions information} is available at \url{www.nature.com/reprints}.

%

\onecolumn
\clearpage

\section*{\LARGE Supplementary Information}
\renewcommand{\figurename}{\hspace{-4pt}}
\renewcommand{\thefigure}{Supplementary Fig.~\arabic{figure}}
\renewcommand{\theHfigure}{Supplementary Fig.~\arabic{figure}}
\renewcommand{\tablename}{\hspace{-4pt}}
\renewcommand{\thetable}{Supplementary Table \arabic{table}}
\renewcommand{\theHtable}{Supplementary Table \arabic{table}}
\setcounter{figure}{0}
\setcounter{table}{0}

\begin{figure*}[!htbp]
    \centering
    \includegraphics[width=\linewidth]{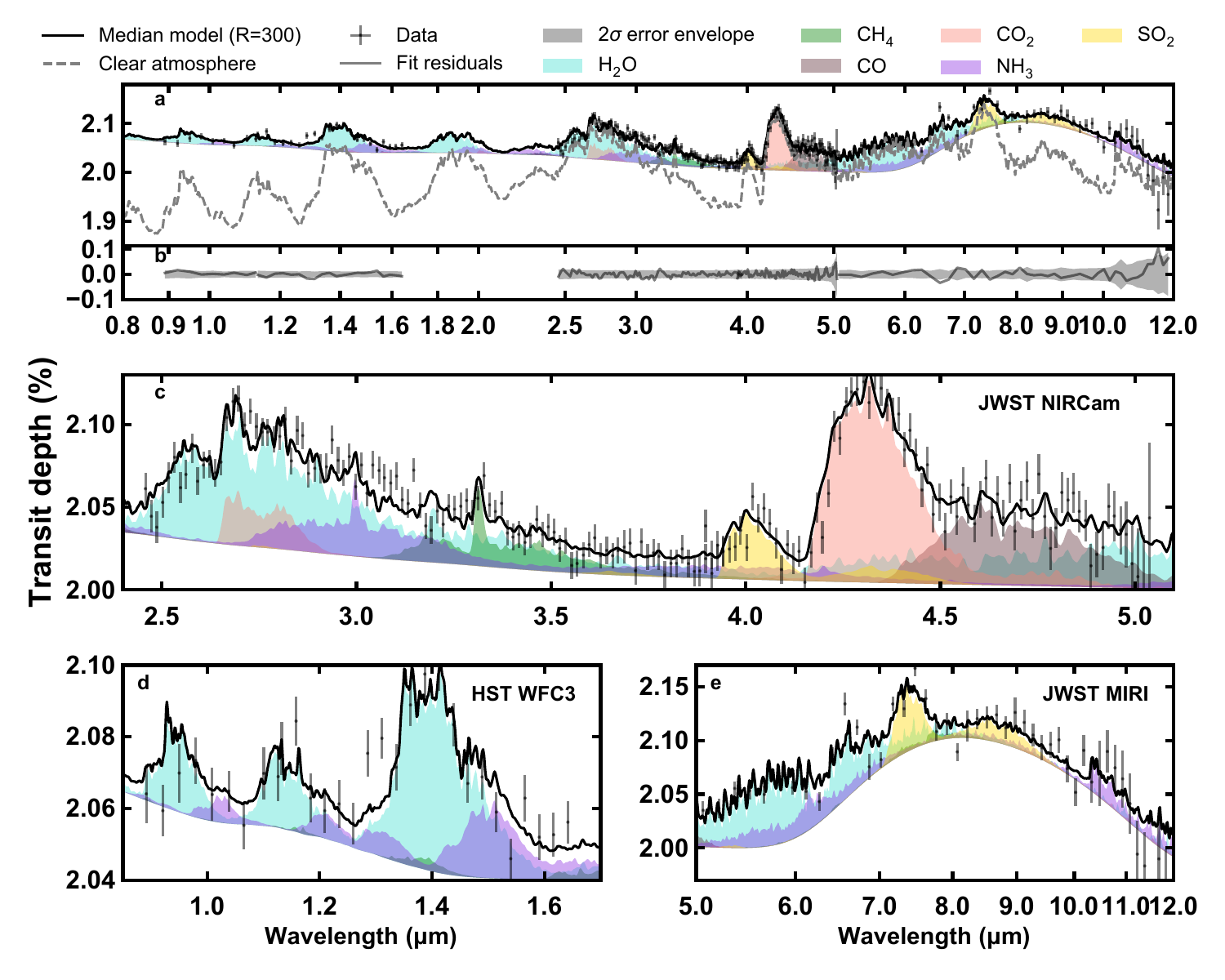}
     \caption{\textbf{Interpretation of WASP-107b's transmission spectrum using free retrievals with Aurora.} The observed transmission spectrum of the planet with 1$\sigma$ error bars is compared to the retrieved transmission spectra using the free-retrieval approach. The retrieved median is shown at a spectral resolution of R=300. The best fit model has a $\chi^2/N_{\rm data}$=1.5. The contributions of the individual detected gases are shown as shaded regions. The gray dashed line shows the clear atmosphere component of the model, that is, the gas contributions without the presence of clouds or hazes. All other elements remain as in Figure 3.}
     \label{fig:spectrum_and_models_free}
\end{figure*}

\begin{figure*}[!htbp]
    \centering
    \includegraphics[width=\linewidth]{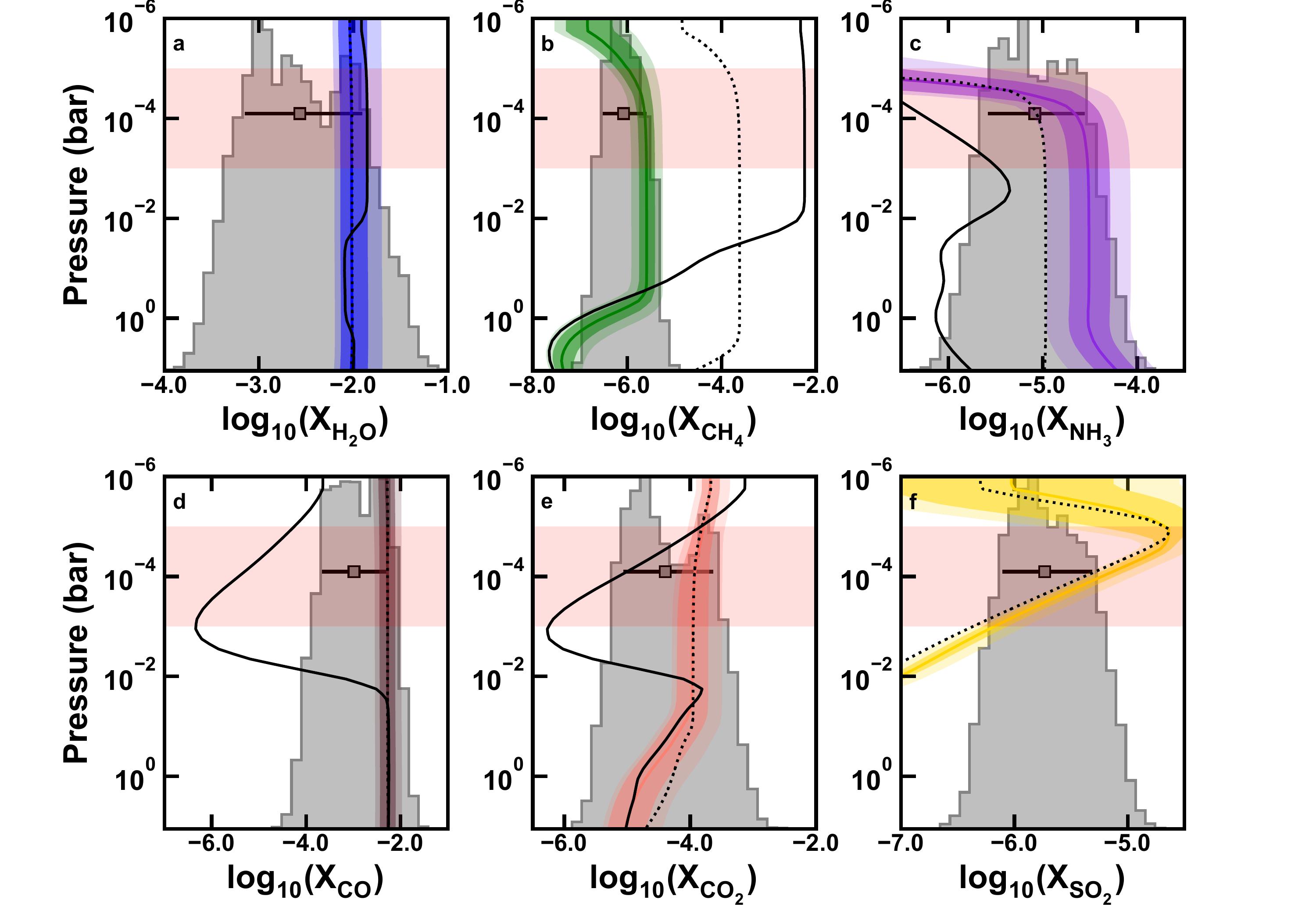}
    \caption{\textbf{Inferred molecular volume mixing ratios from WASP-107b's transmission spectrum with the lower resolution retrievals with Aurora}. Same as Figure 4 but considering the constraints from the free-retrieval with Aurora, with the retrieved volume mixing rations for each detected gas shown in panels \textbf{a}--\textbf{f}. The use of lower resolution (R=20,000) forward models results in constraints that are consistent with those of our fiducial retrieval with CHIMERA but with wider constraints.}
    \label{fig:aurora_constraints}
\end{figure*}

\end{document}